\documentclass[12pt]{article}
\pdfoutput=1
\usepackage{amsfonts} 
\usepackage{graphics}
\usepackage{graphicx}
\usepackage{dblfloatfix} 
\usepackage{fixltx2e}
\usepackage{verbatim} 
\usepackage{amsmath}
\usepackage{multicol}
\usepackage{multirow}
\usepackage{color}
\usepackage{caption}
\usepackage{verbatim}
\usepackage{amssymb}
\usepackage{amsbsy}
\newtheorem{remark}{Remark}

\begin{document}

%==========================================================
%\begin{frontmatter}
  \title{A data porting tool for coupling models with
    different discretization needs}
  \author{S.Ion$^*$, D.Marinescu$^*$, S.G.Cruceanu\footnote{Institute of Statistical Mathematics and Applied
      Mathematics, Romanian Academy, Calea 13 Septembrie
      No. 13, 050711 Bucharest, Romania, Phone/Fax
      (4021)3182439, {\tt email: stelian.ion@ima.ro, dorin.marinescu@ima.ro, stefan.cruceanu@ima.ro}.}, 
   and V.Iordache\footnote{University of Bucharest, Research
      Center for Ecological Services, {\tt email: virgil.iordache@bio.unibuc.ro}}}
\date{June 24, 2014}
      \maketitle

  \begin{abstract}
    The presented work is part of a larger research program
    dealing with developing tools for coupling
    biogeochemical models in contaminated landscapes.  The
    specific objective of this article is to provide the
    researchers a tool to build hexagonal raster using
    information from a rectangular raster data (e.g. GIS
    format), data porting.  This tool involves a
    computational algorithm and an open source software
    (written in C).  The method of extending the reticulated
    functions defined on 2D networks is an essential key of
    this algorithm and can also be used for other purposes
    than data porting.  The algorithm allows one to build
    the hexagonal raster with a cell size independent from
    the geometry of the rectangular raster.  The extended
    function is a bi-cubic spline which can exactly
    reconstruct polynomials up to degree three in each
    variable.  We validate the method by analyzing errors in
    some theoretical case studies followed by other studies
    with real terrain elevation data.  We also introduce and
    briefly present an iterative water routing method and
    use it for validation on a case with concrete terrain
    data.
  \end{abstract}


%==========================================================
\section{Introduction}
\label{intro}
Predicting the long distance effects of local environmental
changes requires a coupling between local and regional
models of ecological and abiotic processes.  Examples
include the integration of local vegetation processes with
regional transport models for heavy metals
(e.g. \cite{Iordacheip, iordachekng, iordachelso} for
reviews and general methodological steps), or the local
resources development with movement of animal species
(e.g. \cite{holmes} for review of Partial Differential
Equations (PDEs)
  used in spatial
interactions and population dynamics, and \cite{gough} for
an application using Geographic Information Systems (GIS)).

Direct coupling of the models is technically possible, but
allows less flexibility for further model
development. Alternatively, more common and flexible
geographical objects are used as an interface among models
of processes occurring at different space-time scales.  One
specific problem with this integration strategy is that the
kind and properties of the geographical objects used as
input to or output from local and regional models are
usually different (see \cite{voinov} for some examples of
problems raised by integrated modelling).  The differences
arise due to methodological constraints related to the
measurements of the space variables and to the modelling
techniques.  Table 1 summarizes the types of geographical
objects which can occur (summarized and developed from
\cite{goodchild}).
\begin{table}[htbp!]
  \caption{Types of geographical objects 
      occurring directly and indirectly in the modelling of
      coupled environmental process (reconstructed and
      adapted from \cite{goodchild}).  Field type approach
      allows a rigorous description of the error and
      empirical verification, while discrete type approach
      does not allow a good treatment of errors, usually
      involves a filtering of empirical data.}
  \centering
  {\scriptsize
    \begin{tabular}
      {p{0.15\linewidth}p{0.23\linewidth}p{0.26\linewidth}p{0.24\linewidth}}
     % {p{0.17\linewidth} p{0.24\linewidth} p{0.24\linewidth} p{0.23\linewidth}}
      \hline
      Approach
      &
      Measurement, observation
      &
      Primary (``real'') geographical objects
      &
      Derived (``methodological'') geographical objects
      (resulted from data modelling and plane discretization)
      \\
      \hline
      Field type (properties with relation of spatial location)
      &
      Variables $z$ - observable at a scale much smaller than
      the derived geographical objects and the empirical
      precision of geographical location
      \medskip

      Variables $z$ - observable at a scale larger than the
      potentially derivable geographical objects and the
      empirical precision of geographical location
      &
      Tuples  $<x, y, z_1,\ldots,z_n>$ with space variables $z$
      and location $(x,y)$.
      \medskip
    
      Tuples with $(x, y)$ on the observation polygon or on
      the transect line (e.g the center of the 50 x 50 m
      plot).
      \medskip

      Interpolated field (the infinite set of tuples).
      &
      Substrates with attributes (polygons, contour lines,
      regularly distributed points as centers of a plane
      discretization).  The polygons are characterized by a
      variation of the variables inside them (a constant in
      the simplest case). The size of the polygons is not
      empirically constrained; 
      \medskip

      Substrates with attributes.  It makes no natural sense
      to have the size of the discretization units smaller
      than the observation scale of the spatial variables.
      \\
      \\
      Discrete type (substrate located in space with
      properties) 
      &
      The observation scale does not influence the approach.
      Usually this approach make use of old geographical maps
      or methodological objects resulted from field approach.
      &
      Points, polygons, lines filling and empty geographical
      space.
      &
      Field obtained by planar enforcement (points with a
      certain value inside the discrete object, by class, and
      a constant value in the empty space).
      \\
      \hline
  \end{tabular}
  }
\end{table}

It can be seen from the inspection of this table that raster
data and vector data are not basic terms in geographical
ontology (see also \cite{ag}).  For instance a raster refers
in principle to a square tessellation of the plane with a
constant value of the space variable function inside each
square.  Although raster data models are not a primary
information source, in practical modelling they are
frequently the only available primary data source scales
(e.g.  in digital terrain models).  A large volume of GIS
raster data is now-a-days collected and used for scientific
research purposes, as well as for different practical
applications, and a wide variety of software has been
developed over time for reading and processing it
\cite{arcgis, rlang} (one may also see \cite{varekamp} for a
practical guide to use public domain geostatistical and GIS
software).  The data of a raster are organized into rows and
columns and structured as a matrix.  Any information of
interest is characterized by an unique value in each cell
which maybe ``null'' if no data is available.  This
information may represent continuous data, as elevation,
temperature, rainfall intensity or discrete (thematic,
nominal) data, as land use or soil category, \cite{arcgis}.
The fidelity of a raster data with respect to the real
information is a challenging issue \cite{epa, mackay, mitas,
  moore, smith} that must be taken into consideration since
the errors propagate on raster data outputs \cite{pelletier,
  vasqez, walker}.  Usually, large cells entail poor
accuracy.  Some mathematical models of environmental
phenomena require information at a subgrid-scale.  For
example, the input data for partial differential equations
must satisfy a certain degree of smoothness.  Also, the
domain discretization may be better suited to configurations
other than that of common square cells.  Such situations may
appear when one uses finite volume or finite element methods
to numerically solve PDEs, as well as when one uses models
based on hexagonal cellular automata \cite{baetens}.  With
the development of laser altimetry in geography and ecology
(\cite{kumh, Srinivasan}), point clouds tend to be more
often used for direct analysis within a GIS.  Hexagonal
lattices can be used for the extraction of knowledge by
clustering techniques from such data (e.g. \cite{jiang,
  Hagenauer}).  Using the rasters obtained by interpolations
of LiDAR data, \cite{Moreira} have pointed out the important
influence of DTM resolution on the estimated soil loss by
erosion modelling. This influence could be explored also by
erosion models using hexagonal lattices if the porting
software would be available.  In this context, one can say
that there is a need to build spatial interpolation
algorithms for the purpose of porting rectangular raster
information to other networks with cells of a different size
or geometry, i.e. a method of {\it data porting} (DP).
These algorithms should be developed in such a way to
preserve as much as possible the original measured space
variables.  In this article we tackle the particular
situation of variables treated by a field type approach,
observable at a scale much smaller than the derived
geographical objects and the empirical precision of
geographical location, having the size of the derived and
not empirically constrained polygons.  In particular, we
present a DP method to construct a hexagonal raster using a
rectangular raster data input. To the best of our knowledge
and according to \cite{wei}, there are currently no
databases provided in ``hexagonal raster format''.
%==========================================================

%==========================================================
\subsection{Data porting methods}
A rough classification divides DP into one stage and
multi-stage methods.  In a one stage method, the data value
corresponding to a cell $c_i$ of a new grid is given by the
direct inspection of the values corresponding to the old
grid cells that intersect $c_i$.  The methods in this class
(e.g. the {\it nearest neighbor}, \cite{dodgson}, {\it area
  weighted mean}, \cite{wei}, {\it and kriging methods},
\cite{kri}) approximate the discrete raster function.  In
\cite{gardner}, the authors develop a rescaling method also
applicable to the transformation of grid configuration (from
rectangular to triangular or hexagonal).  The method is
based on sampling points on the original grid located around
the point in the center of a pixel of the rescaled grid.
The kriging methods are well suited for irregular sample
points and they are not effective on regular and dense
grids.  Moreover, they require a set of information
concerning the correlation functions.  Such information must
be known in advance or can be inferred by a proper process
from the existing data (see also \cite{mat, dub}).  The
multi-stage class methods \cite{dodgson, wolberg} assume the
existence of one or more intermediate (everywhere defined)
functions from which the cell values in the new grid are
sampled.  In this article, we propose a two stage method
similar to one widely used in image resampling,
\cite{dodgson, wolberg}.  The basic assumption is that the
raster values represent the point values of an everywhere
defined function that models a certain physical property.
In most cases, the function is not analytically known and
the raster values are calculated from the values of this
function on a set of irregular spatially distributed points.
The idea of constructing an intermediate stage is that an
extension function is closer than the raster to the model
function.  Using this method, the hexagonal raster tends to
approximate the model function rather than its raster
representation.  The construction of the intermediate
function is essentially based on 2D bicubic interpolation.
The bicubic interpolating polynomial is widely used in
geomorphology, hydrology, image processing, computer
graphics, \cite{mitchel}.  It is known that the way one
chooses the interpolating polynomial is not unique, but it
strongly depends on the specifics of the data one has to
interpolate.
%==========================================================

%==========================================================
\subsection{Applications of hexagonal raster}
As an application for this hexagonal raster, we propose an
iterative cellular automaton based method to model the
drainage network of a given landscape.  Such channel
networks are of interest in domains as hydrology,
geomorphology, ecology etc.

The water flow  on hill slopes is a  complex phenomenon that
needs  a   lot  of   information  concerning   the  physical
properties about the  soils and plant cover  of the terrain.
Focusing only on  water path, one ignores  all these details
and  builds  simplified models  based  only  on the  terrain
topography.  In this approach, the water is viewed as a thin
film  flowing  (like  rolling  balls) from  a  cell  to  its
adjacent neighbors.  There  are at least two  reasons to use
hexagonal instead of rectangular  raster: the cell adjacency
and the local symmetry of  the hexagonal structure.  For the
square grid case, each cell has four adjacent neighbors with
which  shares a  face and  other  four with  which shares  a
corner  (all these  eight neighbors  forming the  well known
Moore neighborhood\footnote{There are situations when models
  with Moore type neighbors  are suitable, e.g. the discrete
  approximation  of  PDEs.  On  the  other  hand, there  are
  physical  situations  where  the boundary  length  between
  cells  and a  better  network symmetry  play an  important
  role,  and  therefore this  type  of  neighborhood is  not
  anymore adequate.}), and thus,  some kind of anisotropy is
induced  by  this  type  of  structure.   In  contrast,  the
adjacent cells in  a hexagonal raster are of  the same type,
all six neighbors sharing an  edge with the central cell and
being at the same distance  from its center; this implies no
ambiguity  in defining  the  first order  neighborhood of  a
given  cell.   Such  kind  of  adjacency  and  the  superior
symmetry  of  the  cells  make  the  hexagonal  raster  more
suitable to model water  flow, erosion, population migration
etc.  By example, it was observed, \cite{deSousa}, that flow
direction vectors are better preserved for hexagonal instead
of  square tilled  grids when  one moves  from one  scale to
another.  Moreover,  concerning the properties  of symmetry,
\cite{Frisch}  showed that  the mean  values obtained  using
square lattice-gas cellular automata  models do not obey the
Navier-Stokes equations  (although the model  conserves mass
and momentum)  due to the  lack of enough  symmetry, whereas
hexagonal   symmetry   is  sufficient   \cite{wolf}.    This
emphasizes  the local  group  of symmetries  can have  major
implications when passing from local to global level.

For regular square networks, there are many methods
\cite{callagan, quinn, seibert, tarboton} to extract the
drainage network.  All these methods assume a flow routing
algorithm and a criterion for establishing if a cell belongs
to or doesn't belong to the discharge network.  The common
drawbacks of such approaches are the drainage pattern's
dependence on the grid resolution \cite{pelletier}, the
problems of pits (there is no downstream flow from a cell
pit) and flat regions (the flow direction cannot reliably be
obtained from the neighboring cells \cite{hutchinson,
  soille}) We propose here a hybrid method which combines a
hexagonal flow routing scheme with cellular automaton.  Due
to their simplicity, reduced numerical effort, and fast
computational speed, cellular automata are currently
frequently used for numerical simulation of natural
phenomena, see \cite{chiba, ambrosio01,fonstad, mortlock,
  molofsky, silva, caesar} to cite a few.  Regular cellulars
based on squares are directly compatible with GIS rasters
and therefore, they are almost always met when modelling
with such input data.  However, there are situations when
simulations by cellular automata on hexagonal grids are more
suitable than the ones on rectangular grids, \cite{birch,
  ambrosio02, fonstad, sahr}.

In Section~\ref{Method} we present a general overview of DP
method we are proposing in this article and we argue in its
favor.  After introducing some necessary notations, we
describe in Section~\ref{sect_2D_cubic_extension} the
Essentially Non-Oscillating (ENO) and Outlier Filtering (OF)
algorithms used to build the 1D interpolant.  We then
present the construction of the 2D intermediate function
which uses the above 1D interpolant.  The last part of this
section is devoted to the process of DP where we first
describe the construction of a hexagonal network and then
present the numerical GIS data conversion of this network.
The algorithms accompanying the DP method are detailed in
\ref{ApA}.  Section~\ref{sect_numerical_applications} is
dedicated to the validation process of DP method and is
divided into two parts.  We analyze the errors of DP using a
theoretical and then a practical dataset consisting of
terrain elevation data provided by GIS rasters.  The second
part of this section describes a routing method built for
the hexagonal raster.  The routing method is based on a
simplified model of hexagonal cellular automaton for water
flow and is applied to real GIS data of some zones from the
Romanian territory.  The conclusions, perspectives and
observations are presented in the last section of this
article.
%==========================================================

%==========================================================
\section{Method}
\label{Method}
Usually, when dealing with exact interpolation by smooth
functions, there is a class of data that introduces spurious
oscillations (Gibbs phenomenon, \cite{lehman}).  In order to
eliminate this phenomenon, we propose an ENO method.  ENO
provides an exact and continuous interpolant for a given 1D
reticulated function, and tries to minimize (in some sense
described in Section~\ref{sect_1D_ENO}) the oscillations
produced by interpolation on the discontinuity zones of the
reticulated function.

Another undesirable complication is related to the possible
presence of the data outliers.  In this case, before using
any interpolation method, one usually tries to filter these
outliers.  In this article, we propose a direct
interpolation OF method. OF eliminates the eventually rare
abnormal data of a reticulated function, but does not always
provide an exact and continuous interpolant.

The transition from the square to a hexagonal raster data is
accomplished in four steps.  First, starting from the square
raster, one defines the 2D reticulated function on the set
of all square centers; its value at any such center is equal
to the raster datum of the square containing that center.
Using this function, one then constructs the intermediate
(extension) function which will be defined at any point from
the square raster domain.  Next, one defines the hexagonal
cellular network, and finally constructs the hexagonal
raster: the value of the hexagonal raster function on any
cell will be given by the value of the intermediate function
at its center.
%==========================================================

%==========================================================
\section{2D Bi-cubic Extension}
\label{sect_2D_cubic_extension}
The function extension problem is too wide a field of
mathematics to be tackled in a single paper, so we restrict
ourselves to a narrow subject that can be formulated as
follows.  Let $\mathcal{D}$ be a rectangular domain in
$\mathbb{R}^2$ and $\mathcal{N}$ a finite set of points
$(x_i,y_j)$ in $\mathcal{D}$.  A reticulated function is
defined by
\begin{equation}
  \label{cubic_ext.02}
  g:\mathcal{N}\to \mathbb{R}, \;
  \mathcal{N}=\bigcup_{i,j} \left\{ (x_i,y_j) \right\}, \;
  g(x_i,y_j)=g_{i,j}.
\end{equation}
If one thinks of the reticulated function $g$ as a
restriction to $\mathcal{N}$ of a function $G$ defined on
the entire domain $\mathcal{D}$, then it makes sense, when
given $g$, to ask about the values of $G$ at a point outside
of $\mathcal{N}$ (see \cite{stasch} for an interesting
discussion concerning the meaningfulness of spatio-temporal
interpolation of the discrete data).

By extension of the reticulated function
(\ref{cubic_ext.02}) one means a numerical scheme to
estimate the value of $G$ at any point
$(x,y)\in\mathcal{D}$.  A bi-cubic extension of the
reticulated function (\ref{cubic_ext.02}) is a function
defined on the entire domain $\mathcal{D}$ that is a
piecewise bi-cubic polynomial function
$\widetilde{g}\in\pi_{3,3}$, where $\pi_{3,3}$ designates
the set of all polynomials of degree three in each of the
two variables $x$ and $y$
\begin{equation}
  \label{cubic_ext.03}
  \widetilde{g}:\mathcal{D}\to \mathbb{R}, \;
  \mathcal{D}=\bigcup_a\omega_a, \;
  \widetilde{g}(x,y) \Big| _{\omega_a}\in\pi_{3,3},
\end{equation}
with $\omega_a$ the elements of a cell partition of
$\mathcal{D}$.

The problem of the bi-cubic extension can be formulated as
follows: {\it given a reticulated function {\rm
    (\ref{cubic_ext.02})}, define the extension {\rm
    (\ref{cubic_ext.03})} that approximates the model
  function $G:\mathcal{D}\to \mathbb{R}$}.

In what follows, we propose two methods to define a bi-cubic
extension, both of them assuming a 1D extension method.  The
2D extension method is obtained by successively applying the
1D method once along the $Ox$ and then along the $Oy$
directions.

To define the 1D algorithm, we need to introduce some
notations.  Let $\Xi:=\{\xi_i\}_{i=\overline{1,N}}$ be a set
of 1D consecutive knots inside a generic interval
$[\alpha,\beta]$, with
$\alpha\leq\xi_1<\ldots<\xi_N\leq\beta$, and
$\{f_i\}_{i=\overline{1,N}}$ the values of some reticulated
function $f$ at these points.  There are different formula
to write the unique cubic polynomial that equals the values
$f_{i_k}$ at four distinct knots
$\{\xi_{i_k}\}_{k=\overline{1,4}}$.  The Newton's form of
the Lagrange interpolation polynomial reads as \cite{Sc}
\begin{equation}
  \label{cubic_ext.04}
  \begin{array}{ll}
    \mathcal{P}_{(i_1, i_2)}^{f;(i_3,i_4)}(\xi)&:=
    f(\xi_{i_1})+(\xi-\xi_{i_1})[\xi_{i_1};\xi_{i_2}]f+\\
     &+(\xi-\xi_{i_1})(\xi-\xi_{i_2})\Big(
    [\xi_{i_1};\xi_{i_2};\xi_{i_3}]f+  
     (\xi-\xi_{i_3})[\xi_{i_1};\xi_{i_2};\xi_{i_3};\xi_{i_4}]f)\Big),
  \end{array}
\end{equation}
where [$\cdot $;$\cdot $], [$\cdot $;$\cdot $;$\cdot $] and
[$\cdot $;$\cdot $;$ \cdot $;$\cdot $] represent the divided
difference operators of order one, two and three,
respectively. The divided difference operator is recursively
defined as
\begin{equation*}
  \begin{aligned}
    \left[\xi_{i_1};\xi_{i_2}\right]f & = \displaystyle\frac{f_{i_2}-f_{i_1}}{\xi_{i_2}-\xi_{i_1}},\\
    \left[\xi_{i_1}; \ldots ;\xi_{i_{n+1}}\right]f & =
    \displaystyle\frac{\left[\xi_{i_2};\ldots
        ;\xi_{i_{n+1}}\right]f-\left[\xi_{i_1};\ldots
        ;\xi_{i_n}\right]f}{\xi_{i_{n+1}}-\xi_{i_1}},
  \end{aligned}
\end{equation*}
where the knots $\xi_{i_j}$ are all different from each
other.

For any $k=\overline{1,N-1}$, we denote by $I_{k}^{\xi}$ the
open interval
\begin{equation}
  \label{Ik}
  I_{k}^{\xi} := (\xi_k,\xi_{k+1}),
\end{equation}
and by $S_{k}^{\xi}$ the set of up to six consecutive knots
\begin{equation}
  \label{Sk}
  S_{k}^{\xi} := \left\{ \xi_{k-2},\xi_{k-1},\xi_{k},
    \xi_{k+1}, \xi_{k+2},\xi_{k+3} \right\} \cap \Xi.
\end{equation}
$S_{k}^{\xi}$ defines a neighborhood of knots that can
influence the values of the extension function of $f$ on the
interval $I_{k}^{\xi}$.
%==========================================================

%==========================================================
\subsection{1D Essentially Non-Oscillating Extension (ENO)
  Algorithm}
\label{sect_1D_ENO}
The ENO type extension operator was first introduced in the
context of numerical approximation of hyperbolic system of
partial differential equations, \cite{harten1, harten2}, in
order to reduce the Gibbs oscillations that appear in the
reconstruction of the shock wave solution.  In \cite{SD},
the ENO algorithm (\ref{cubic_ext.05}) was used to perform
the multiresolution analysis of a 1D reticulated function
defined on a bounded interval.  For each $I_{k}^{\xi}$, the
method defines an interpolant
$\mathcal{P}_{(i_1,i_2)}^{f;(i_3,i_4)}(\xi)$ associated to
the four distinct consecutive knots $\xi_{i_1}, \xi_{i_2},
\xi_{i_3}, \xi_{i_4}$ from $S_k^{\xi}$, by setting $i_1:=k$,
$i_2:=k+1$, and choosing $i_3$, $i_4$ such that
$\mathcal{P}_{(k,k+1)}^{f;(i_3,i_4)}(\xi)$ has, in some
sense, the smallest oscillation on the interval
$\bar{I_{k}^{\xi}}$.

Specifically, we compare the above cubic polynomials
associated to the interval $\bar{I_{k}^{\xi}}$ with the
linear interpolant corresponding to the same interval
because linear interpolation is exact and it does not have
oscillations.  We want to select the ``nearest'' cubic
polynomial with respect to the linear interpolant. In order
to achieve this, we choose as a measure of oscillation, the
$\mathbb{L}^2(I_k)$ distance\footnote{We choose the
  $\mathbb{L}^2$ distance for attenuating interpolation
  oscillations because the resulting formulas are quite
  simple and easily to introduce in the code. One can use
  $\mathbb{L}^1$ or some $\mathbb{L}^p$ distance.  We have
  no arguments that one of them is the best.} between
$\mathcal{P}_{(k,k+1)}^{f;(i_3,i_4)}$ and the linear
interpolant
$\mathcal{Q}_{k}^{f}(\xi):=f(\xi_k)+(\xi-\xi_k)[\xi_k;\xi_{k+1}]f$.
Therefore, using the simplifying notations
$\mathcal{P}_{-1}^{f}:=\mathcal{P}_{(k,k+1)}^{f;(k-2,k-1)}$,
$\mathcal{P}_{0}^{f}:=\mathcal{P}_{(k,k+1)}^{f;(k-1,k+2)}$,
$\mathcal{P}_{1}^{f}:=\mathcal{P}_{(k,k+1)}^{f;(k+2,k+3)}$,
the 1D ENO algorithm can now be sketched as
\\
\begin{minipage}{\linewidth}
  \vspace*{1em} \centering \hrule
  \begin{equation}
    \begin{array}{l}
      \textnormal{Find } a \textnormal{ s.t. }
      a = 
      {
        \operatorname*{\arg\,\min}\limits_{b\in\{-1,0,1\}}
      } 
      \left\|\mathcal{P}_{b}^{f} - \mathcal{Q}^{f}_k
      \right\|_{\mathbb{L}^{2}(I_{k}^{\xi})}.\\
      \textnormal{If } a \textnormal{ is        not unique,
        then choose } a:=0.\\
        \textnormal{Set }
        \widetilde{f}(\xi):=\mathcal{P}_{a}^{f}(\xi),\quad
        \forall \xi\in \bar{I_{k}^{\xi}}:=[\xi_k,\xi_{k+1}].
    \end{array}
    \label{cubic_ext.05}
  \end{equation}
  \hrule \medskip
\end{minipage}
%==========================================================

%==========================================================
\subsection{1D Outlier Filtering Extension (OF) Algorithm}
\label{sect_1D_OF}
As in the ENO algorithm, for each $I_{k}^{\xi}$, one
determines the polynomial
$\mathcal{P}_{(i_1,i_2)}^{f;(i_3,i_4)}$ such that the
quantity
\begin{equation*}
  \left|\left |
    \displaystyle\frac{{\rm d^2}}{{\rm d}\xi^2}
    \mathcal{P}_{(i_1,i_2)}^{f;(i_3,i_4)}
  \right|\right |_{\mathbb{L}^2(I^\xi_k)},
\end{equation*}
is minimized.  Unlike the ENO method, the OF method
generally defines a non-continuous extension function
(possible discontinuities at the knots $\xi_j$).  The
advantage of the method is that the extension function does
not take into account the ``bad'' points (outliers), points
where the function (its value) strongly deviates from a
``regular behavior''.

Both the ENO and OF methods have the property of recovering
polynomial functions of degree up to three, i.e. if the
reticulated function $f$ is generated by a third degree
polynomial $P(\xi)$ then $\widetilde{f}(\xi)\equiv P(\xi)$,
for any $\xi$.

The 1D OF algorithm can now be sketched as:
\\
\begin{minipage}{\linewidth}
  \centering \vspace*{1em} \hrule
  \begin{equation}
    \label{cubic_ext.06}
    \begin{array}{l}
      \textnormal{Find } i_1,i_2,i_3,i_4 \textnormal{ s.t. }\\
      \left\{\xi_{i_l}\right\}_{l=\overline{1,4}} = 
      {
        \operatorname*{\arg\,\min}
        \limits_{\xi_l,\xi_m,\xi_p,\xi_q\in S_k^{\xi}}
      } 
      \left|\left |\displaystyle\frac{{\rm d^2}}{{\rm d}\xi^2}
          \mathcal{P}_{(l,m)}^{f;(p,q)}
        \right|\right  |_{\mathbb{L}^2(I^\xi_k)}.\\
      \textnormal{Set } \widetilde{f}(\xi)
        :=\mathcal{P}_{(i_1,i_2)}^{f;(i_3,i_4)}(\xi)\quad
        \forall \xi\in I_{k}^{\xi}.
    \end{array}
  \end{equation}
  \hrule \medskip
\end{minipage}
As previously mentioned, choose the value of $\widetilde{f}$
at any knot $\xi_j$ to be one of the lateral limits or their
average.

Two illustrative examples of 1D ENO and OF methods can be
found in Figures~\ref{teo_eno} and \ref{fig_sin}.
%==========================================================

%==========================================================
\subsection{2D Extension Scheme}
\label{sect_2D_ExtensionAlgorithm}
Let $g$ be a 2D reticulated function of the form
(\ref{cubic_ext.02}).  As the 1D case, one introduces
$I_k^x:=(x_k,x_{k+1})$, $I_l^y:=(y_l,y_{l+1})$ and the
corresponding sets $S^x_k$, $S^y_l$, respectively, of form
(\ref{Sk}).  Denote also by $\omega_{k,l}$ the domain
\begin{equation*}
  \omega_{k,l}=I_k^x\times I^y_l.
\end{equation*}
Let $(x,y)\in \omega_{k,l}$.  The key idea is to define an
extension function as a cross combination of 1D algorithms.
At first, for any $y_m$, one can define an extension
function $\widetilde{f}(x,y_m)$ by using a 1D method with
respect to an $x$ variable.  The restriction
$\widetilde{f}_k(x,y_m)$ of this function to the interval
$I_k^x$ can be written as
\begin{equation*}
  \begin{array}{ll}
    \widetilde{f}_k(x,y_m):=&
    g(x^m_{i_1},y_m)+
    (x-x^m_{i_1})[x^m_{i_1};x^m_{i_2}]g(\cdot,y_m)+\\
       %    (x-x^m_{i_2})\Big([x^m_{i_1};x^m_{i_2};x^m_{i_3}]g(\cdot,y_m)\Big)+\\
    &+(x-x^m_{i_1})(x-x^m_{i_2})[x^m_{i_1};x^m_{i_2};x^m_{i_3}]g(\cdot,y_m)+\\ 
    &+(x-x^m_{i_1})(x-x^m_{i_2})(x-x^m_{i_3})[x^m_{i_1};x^m_{i_2};x^m_{i_3};x^m_{i_4}]g(\cdot,y_m),
  \end{array}
\end{equation*}
where the knots $\{x^m_{i_l}\}_{l=\overline{1,4}}$ belong to
$S_k^x$ and they are given by the 1D method.  By
$[x_{i_1}^m;\ldots;x_{i_n}^m] g(\cdot,y_m)$ one means the
divided difference operator that acts on the reticulated
function $f:=g(\cdot,y_m)$ and the knots
$\{x_{i_j}^m\}_{j=\overline{1,n}}$.

Then, keeping $x$ and having $\widetilde{f}_k(x,y_m)$ for
each $y_m\in S_l^y$, and using again a 1D method, this time
with respect to $y$ variable, one sets the value of the
extension function $\widetilde{g}$ as
\begin{equation*}
  \begin{array}{ll}
    \widetilde{g}(x,y):=&
    \widetilde{f}_k(x,y_{i_1})+ (y-y_{i_1})[y_{i_1};y_{i_2}]\widetilde{f}_k(x,\cdot)+\\ 
     &+(y-y_{i_1})(y-y_{i_2})[y_{i_1};y_{i_2};y_{i_3}]\widetilde{f}_k(x,\cdot)+ \\
     &+(y-y_{i_1})(y-y_{i_2}) (y-y_{i_3})[y_{i_1};y_{i_2};y_{i_3};y_{i_4}]\widetilde{f}_k(x,\cdot).
  \end{array}
\end{equation*}

One notes that the extension function $\widetilde{g}(x,y)$
can have some discontinuous line $x=\overline{x}$ even
inside of a rectangle $\omega_{k,l}$.

See \ref{ApA} for comments on the mathematical properties on
the extension functions $\widetilde{g}$.

When compared to the filter reconstruction, this nonlinear
method is time consuming.  Though, since it is used only
once in the beginning when we transfer data from the GIS
raster to the hexagonal raster, we can benefit by applying
this method since the Gibbs phenomenon is no longer present.
This can be important when big jumps appear in the data (for
example, the landscape elevation in rough terrain).
%==========================================================

%==========================================================
\subsection{Data Porting}
\label{sect_porting_data}
In this section, we present a method to construct the data
set on a regular hexagonal raster starting from a raster
data set used by GIS.  The data transformation from a GIS
raster to a hexagonal raster involves two main steps: the
hexagonal tessellation of the domain on which the GIS raster
is defined and the method of assigning values to the
hexagonal raster cells.  First, we present the geometry of a
hexagonal raster including the number of cells, their size
and relative positions.  Then, we present the method of data
transfer for real valued functions.
%==========================================================

%==========================================================
\subsubsection{Hexagonal raster} 
\label{section_Hexagonal_raster}
The hexagonal raster differs from a GIS raster by the cell
type used to tessellate the area of interest, with regular
hexagons instead of square cells being used for this
tessellation.

The data set can be structured as rows and columns such that
each row has an equal number of cells.  Let $M$ and $N$ be
the number of rows and columns, respectively, and let $r$ be
the radius of the circumscribed circle of the regular
hexagon.  From a computational point of view, besides the
``parameters'' $M$, $N$ and $r$, one needs to know the
position of the hexagon centers with respect to a coordinate
system.

\begin{figure}
  \centering
  \includegraphics[width=0.5\linewidth]{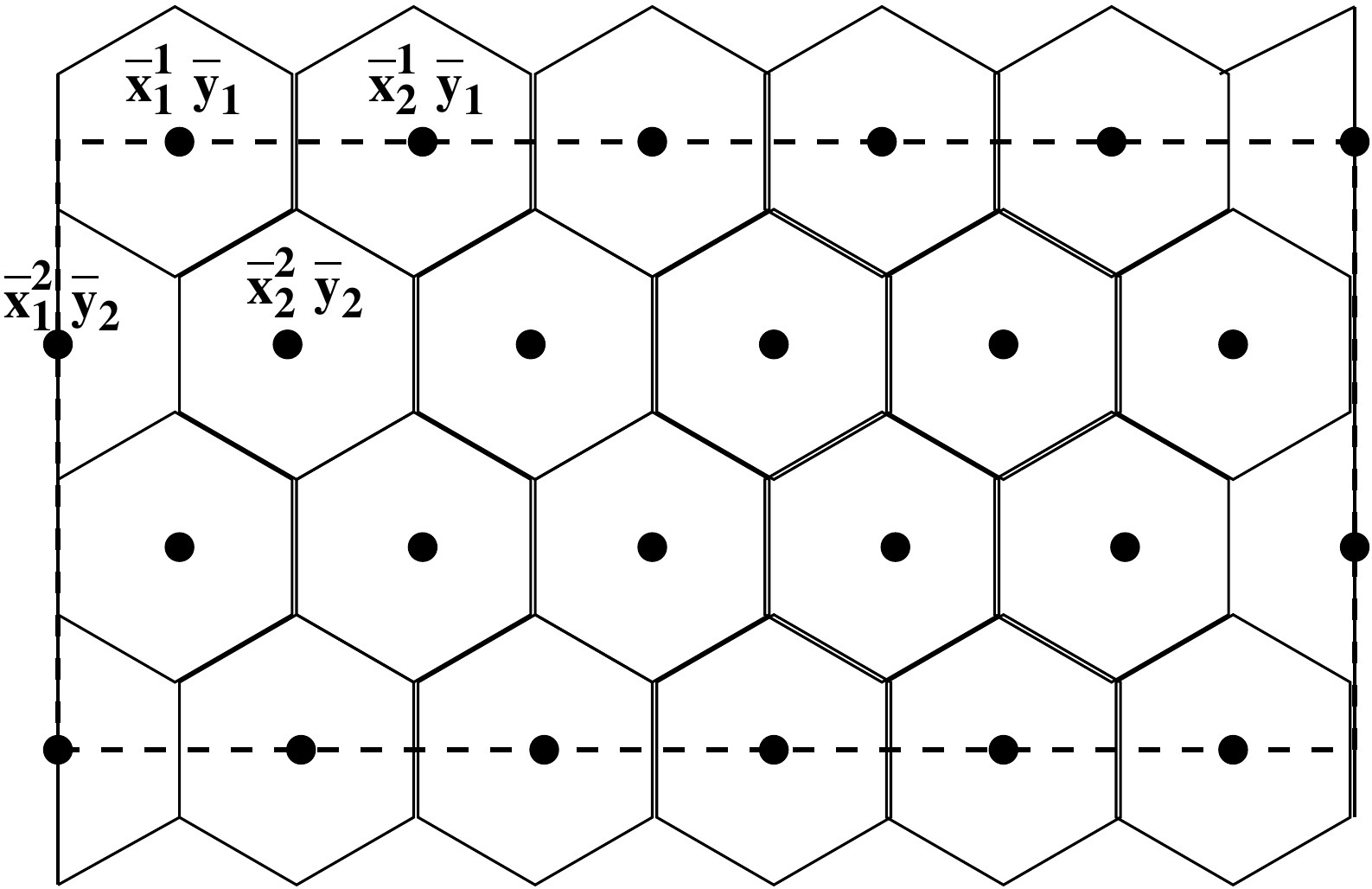}
  \caption{An example of hexagonal tessellation.}
  \label{fig_hexa_tessellation}
\end{figure}

Let ${\mathcal D}$ be a rectangular domain whose sides are
parallel to the $xOy$ coordinate system axes.  A
tessellation of ${\mathcal D}$
(Figure~\ref{fig_hexa_tessellation}) can be defined as
follows.  For any $i=\overline{1,N}$ and $j=\overline{1,M}$,
let $x^h_{i,j}$ and $y^h_{i,j}$ be the coordinates of the
center of the hexagonal cell $h_{i,j}$.  We denote by
$\widetilde{x}_0$ and $\widetilde{y}_0$ the coordinates of
the upper left cell center in the tessellation, and define
the knots $(\widetilde{x}_i^j, \widetilde{y}_j)$ by
\begin{equation*}
  \label{sect4.1_01}
  \begin{array}{ll}
    \widetilde{x}_i^j=\widetilde{x}_0+(i-1)r\sqrt{3}-(j-1)\%2\displaystyle\frac{r\sqrt{3}}{2}, \\ 
    \widetilde{y}_j=\widetilde{y}_0+3(j-1)\displaystyle\frac{r}{2},
  \end{array}
\end{equation*} 
where $a\%2$ represents the remainder of the division of $a$
by $2$.  One now sets
\begin{equation*}
  \label{sect4.1_02}
  x^h_{i,j}:=\widetilde{x}^j_i,\;\; y^h_{i,j}:=\widetilde{y}_j.
\end{equation*}
%==========================================================

%==========================================================
\subsubsection{Porting real numerical data from a
  rectangular to hexagonal raster}
\label{Prnd_section}
The method of a rectangular raster data storage consists of:
a rectangular domain $\mathcal{D}$, a function $G$ defined
on $\mathcal{D}$, a tessellation of $\mathcal{D}$ with
regular rectangular cells $c_{i,j}$, and an approximation
$g^r$ of the function $G$ by constant values $g^r_{i,j}$ on
$c_{i,j}$,
\begin{equation}
  \label{cubic_ext.01}
  g^r:\mathcal{D}\to \mathbb{R}, \;
  \mathcal{D}=\bigcup_{i,j} c_{i,j}, \;
  g^r(x,y) \Big| _{c_{i,j}}=g^r_{i,j}.
\end{equation}

The basic idea is to use an intermediate extension function
in the conversion procedure.  Let $(x_i,y_j)$ be the center
of the cell $c_{i,j}$, $\mathcal{N}$ the set of all these
centers and let $g$ be the reticulated function defined by
\begin{equation*}
  % \label{cubic_ext.02}
  g:\mathcal{N}\to \mathbb{R}, \;
  g(x_i,y_j)=g^r_{i,j}.
\end{equation*}
Using  a 2D extension algorithm, one can construct the
extension function
\begin{equation*}
  \widetilde{g}:\widetilde{\mathcal{D}}\rightarrow\mathbb{R},
\end{equation*}
of $g$ and then define
\begin{equation*}
  g^h:\widetilde{\mathcal{D}}\to \mathbb{R}, \;
  g^h(x,y) \Big| _{h_{i,j}} =
  \widetilde{g}(x^h_{i,j},y^h_{i,j}),
\end{equation*}
where $h_{i,j}$ denotes the hexagonal cell of center
$(x^h_{i,j},y^h_{i,j})$, and
$\widetilde{\mathcal{D}}:=\displaystyle\bigcup_{i,j}h_{i,j}$.
%==========================================================

%==========================================================
\section{Validation and Numerical Applications}
\label{sect_numerical_applications}
\begin{figure}[htbp!]
  \centering
  \begin{tabular}{cc}
  \includegraphics[width=0.5\linewidth]{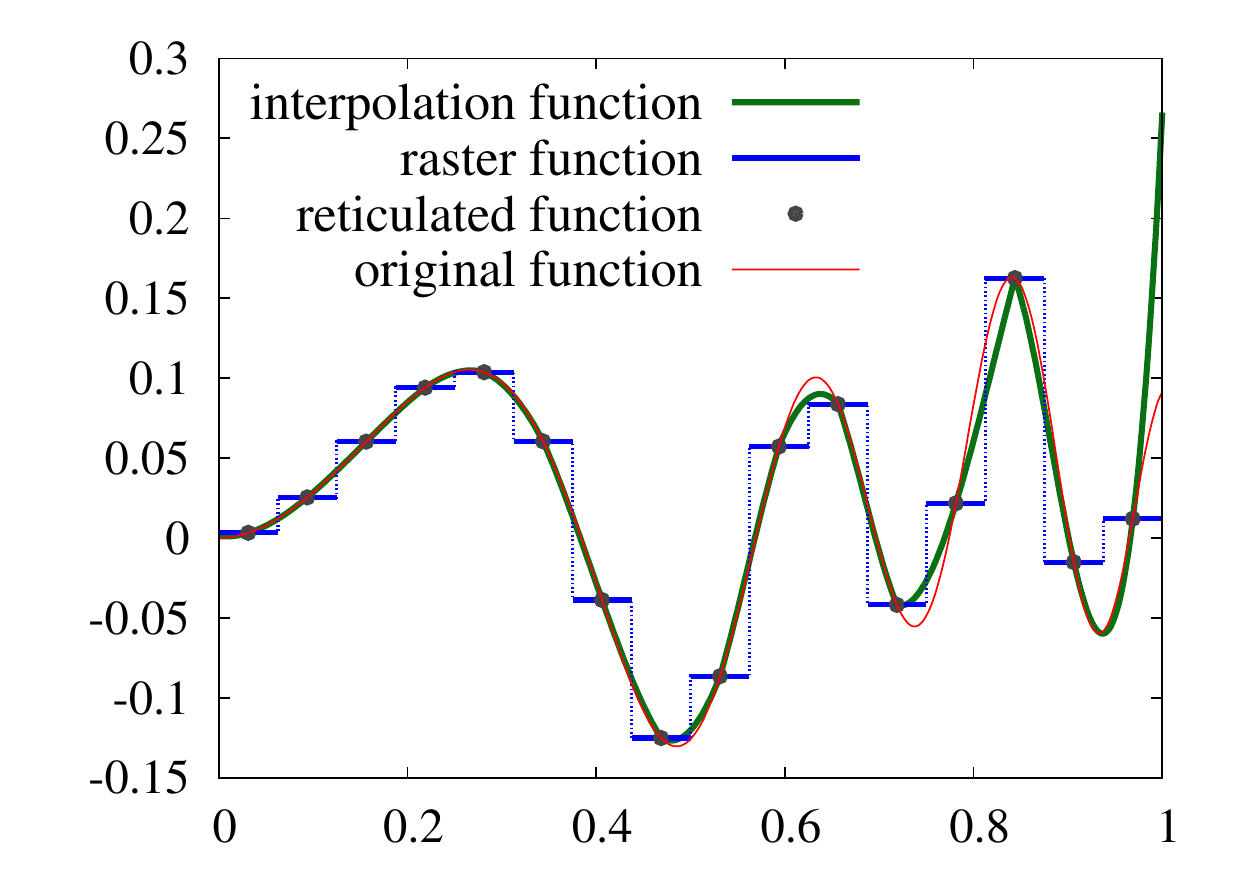}
 \end{tabular}
 \caption{Illustration of the 1D ENO algorithm.  One can
   remark that the interpolated function is ``closer'' to
   the original function (some smooth function which plays
   here the role of the function $G$) than the raster
   function.  Note that {\it it is generally better to
     approximate the original function by the interpolation
     rather than the raster function}.}
  \label{teo_eno}
\end{figure}

\begin{figure}[htbp!]
  \centering
  \begin{tabular}{cc}
    \includegraphics[width=0.45\linewidth]{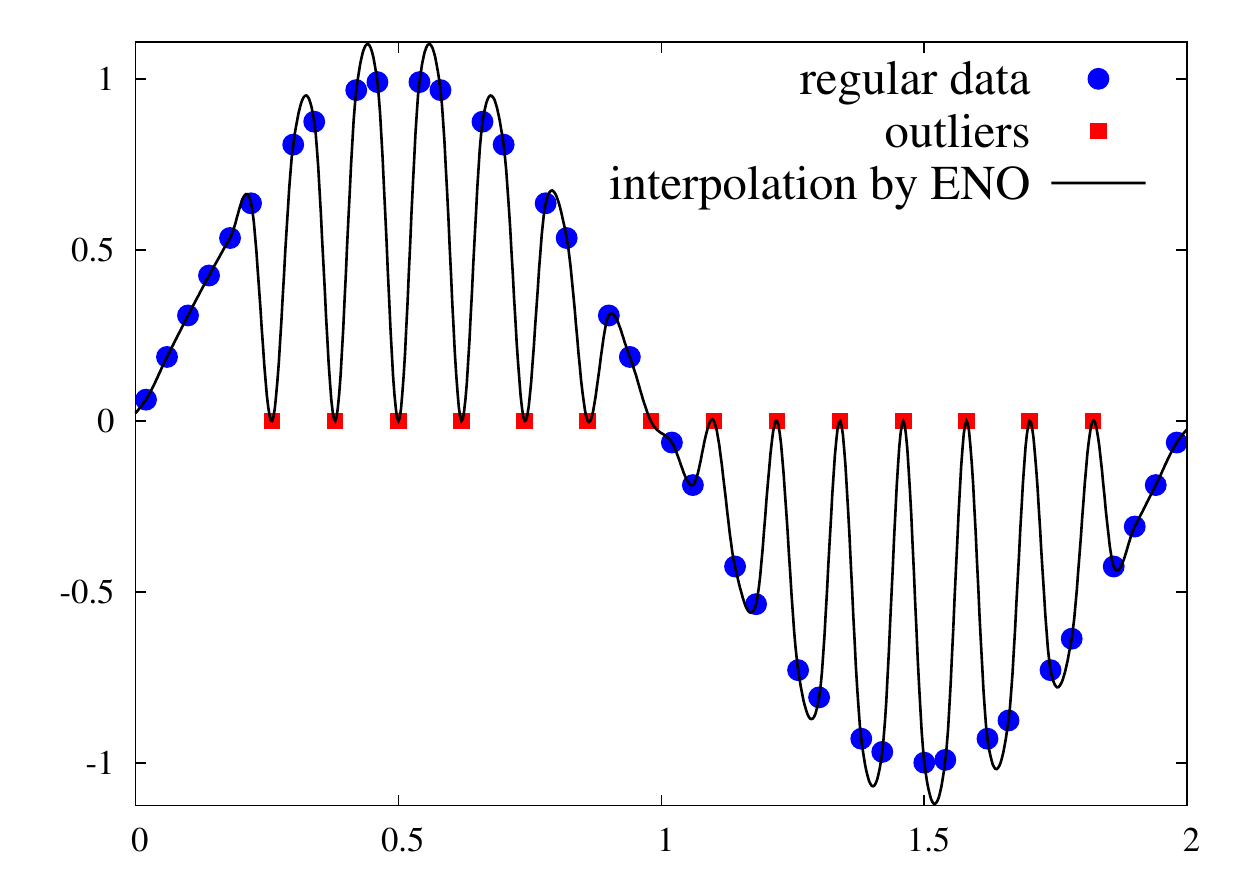}&
    \includegraphics[width=0.45\linewidth]{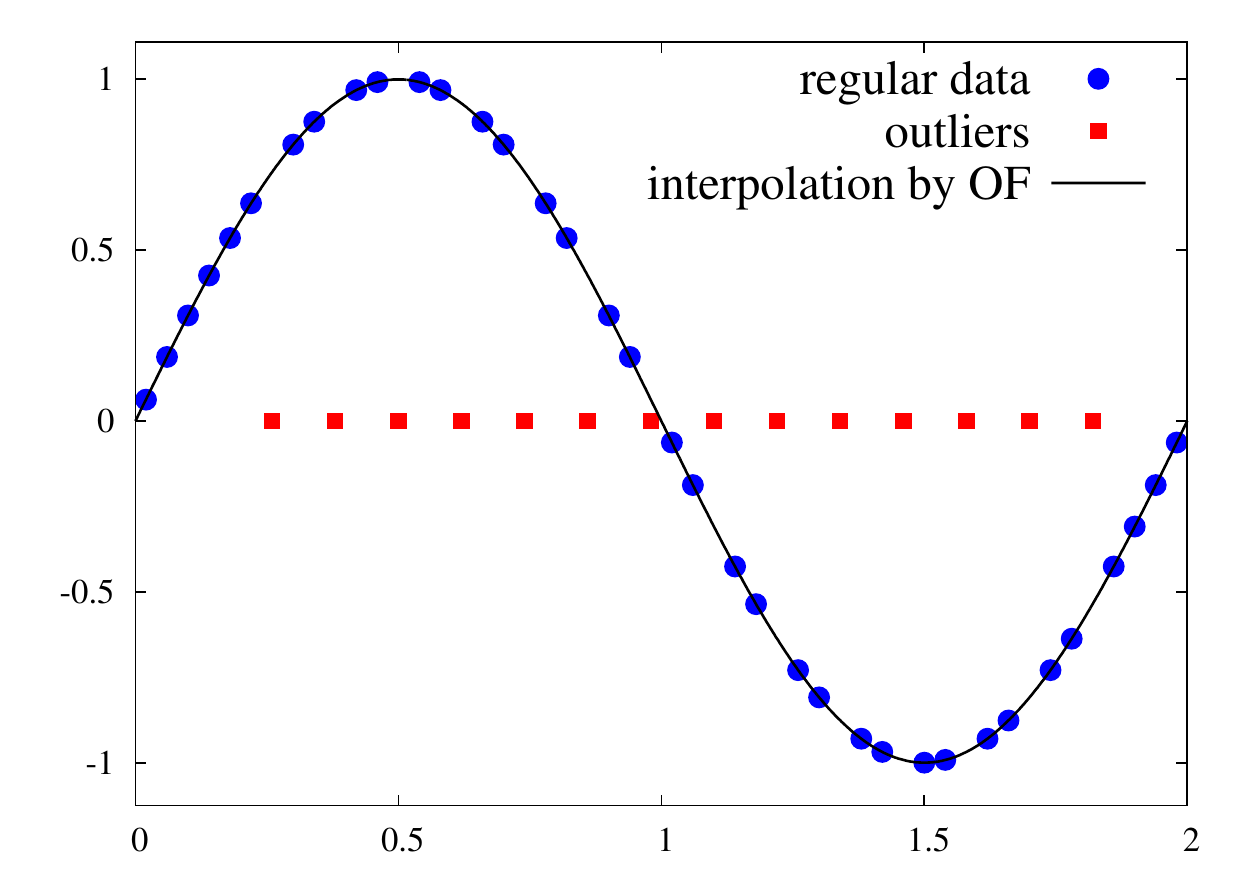}
 \end{tabular}
 \caption{ENO vs OF in the presence of outliers.  Starting
   from $\sin(\pi x)$ we have generated a reticulated
   function where a part of the data have been ``corrupted''
   with the value $0$.  The above figures represent the
   reticulated function (dots) and extension function
   (continuous line) obtained by the interpolation methods
   ENO (on left) and OF (on right).  One can observe that
   the first method is not appropriate for this kind of
   data, it does not recognize the outliers, the second
   method makes a good attempt at this.  This behavior of
   ENO is typical of all exactly interpolating methods.}
  \label{fig_sin}
\end{figure}

The main steps of DP method introduced in
  Section \ref{Prnd_section} can be summarized by the
  following chain
\begin{equation*}
  g^r \rightarrow \widetilde{g} \rightarrow g^h.
\end{equation*}
The accuracy of $g^h$ can be analyzed with respect to the
raster function $g^r$, and with respect to the function, say
$G$, which models the physical property represented by the
raster data.  From an application point of view, the most
important is the accuracy of $g^h$ with respect to $G$.  One
can estimate the errors between $g^h$ and $G$ in some simple
cases (e.g. when $g^r$ is directly obtained from $G$ using
some mathematical operations), but not in general because it
depends on the way $g^r$ is produced.  We now proceed to
analyze the accuracy of our DP method by using some
theoretical examples as well as a case of digital terrain
model.  We begin this section illustrating how the extension
methods work for the 1D case.  Figure~\ref{teo_eno} presents
four graphics: the model function $G$, the raster $g^r$, the
reticulated function, and the extension function
$\widetilde{g}$ obtained by ENO method.  The OF method (when
some outliers are present) is exemplified in
Figure~\ref{fig_sin}.

%----------------------------------------------------------
\subsection{Errors in data porting}
\label{Section_errors_in_DP}
There are four major sources of errors affecting data in a
hexagonal raster obtained by DP (from a rectangular raster)
as a quantitative model of some physical property:

$\bullet$ the error between the values of the physical
property and the mathematical function $G$ which models this
property (the modelling error);

$\bullet$ the error between the function $g^r$ provided by
the GIS raster data and $G$;

$\bullet$ the error between $g^r$ and its interpolant
$\widetilde{g}$;

$\bullet$ the error introduced by the discretization of
$\widetilde{g}$ on the hexagonal cells represented by $g^h$.

As our entry data in DP are raster data, we can only control
the last two components of the error.  The better control of
these errors gives a better representation of $G$ by $g^h$.
We first investigate the accuracy of $g^h$ with respect to
$G$ by considering two theoretical examples where we know
the analytic form of $G$, as well as the way $g^r$ is
obtained from $G$.  We then analyze the errors for some
Digital Elevation Model (DEM).
%----------------------------------------------------------

%----------------------------------------------------------
\subsubsection{Theoretical case study}
Let {\it real physical data} be modeled by the (Runge type)
function of parameter $a$,
\begin{equation}
  G(x,y;a)=\displaystyle\frac{a}{(1+x^2)(1+y^2)},
  \label{eq_G_function}
\end{equation}
and let $g^r$ be a square raster approximation of it.  In
our example, we choose a domain
$\mathcal{D}=[-20,20]\times[-20,20]$ and define two raster
data with regular partitions.  The first one, denoted by
SR1, has $41\times41$ cells, and the second one, denoted by
SR2, has $201\times201$ cells.  For each of these two
partitions, let $g^r$ be a square raster approximation of
$G$ (on any square cell $g^r$ equals the value of $G$ at its
center).

\begin{figure}[htbp!]
  \centering
  \includegraphics[width=0.49\linewidth]{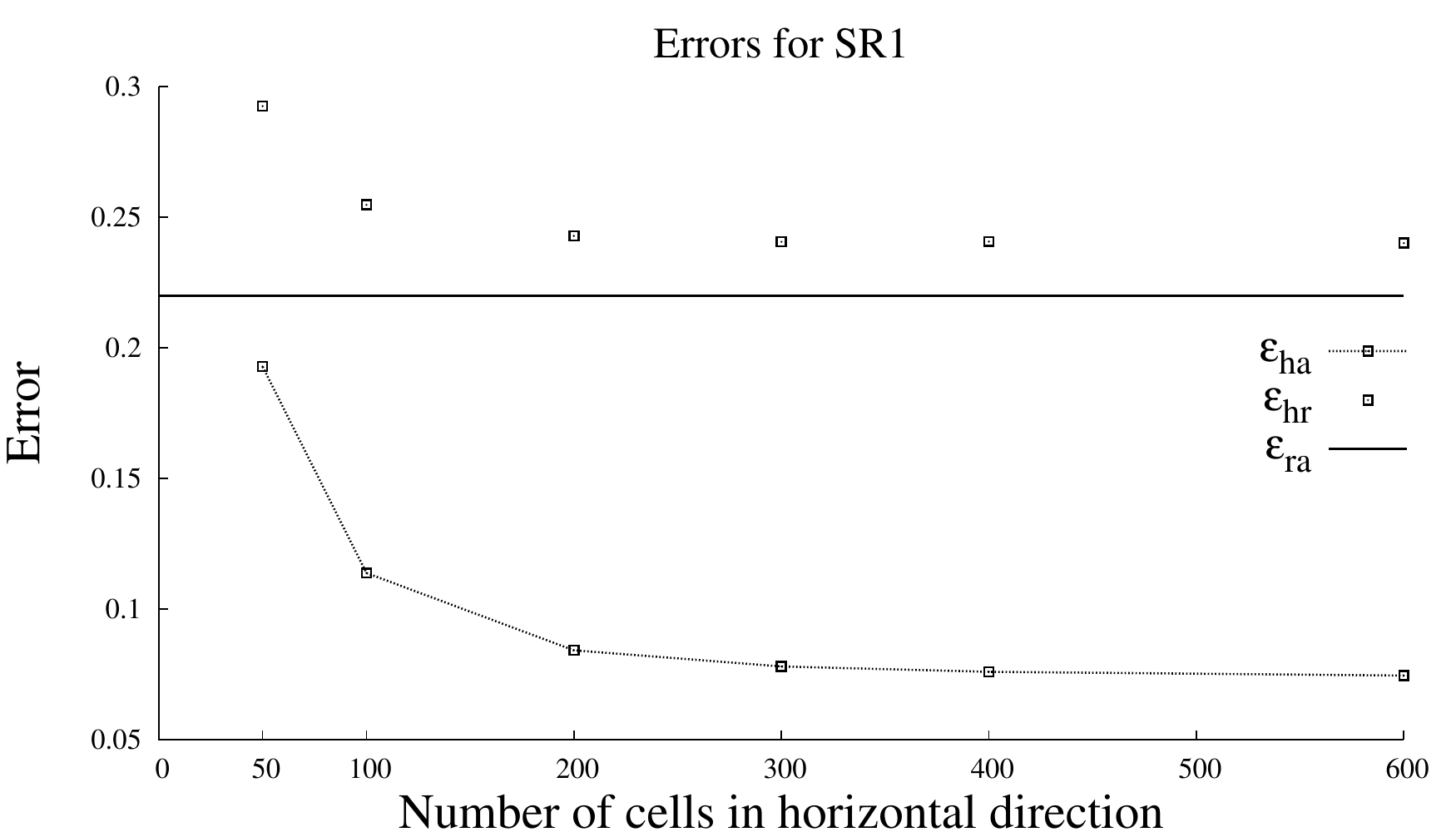}
  \includegraphics[width=0.49\linewidth]{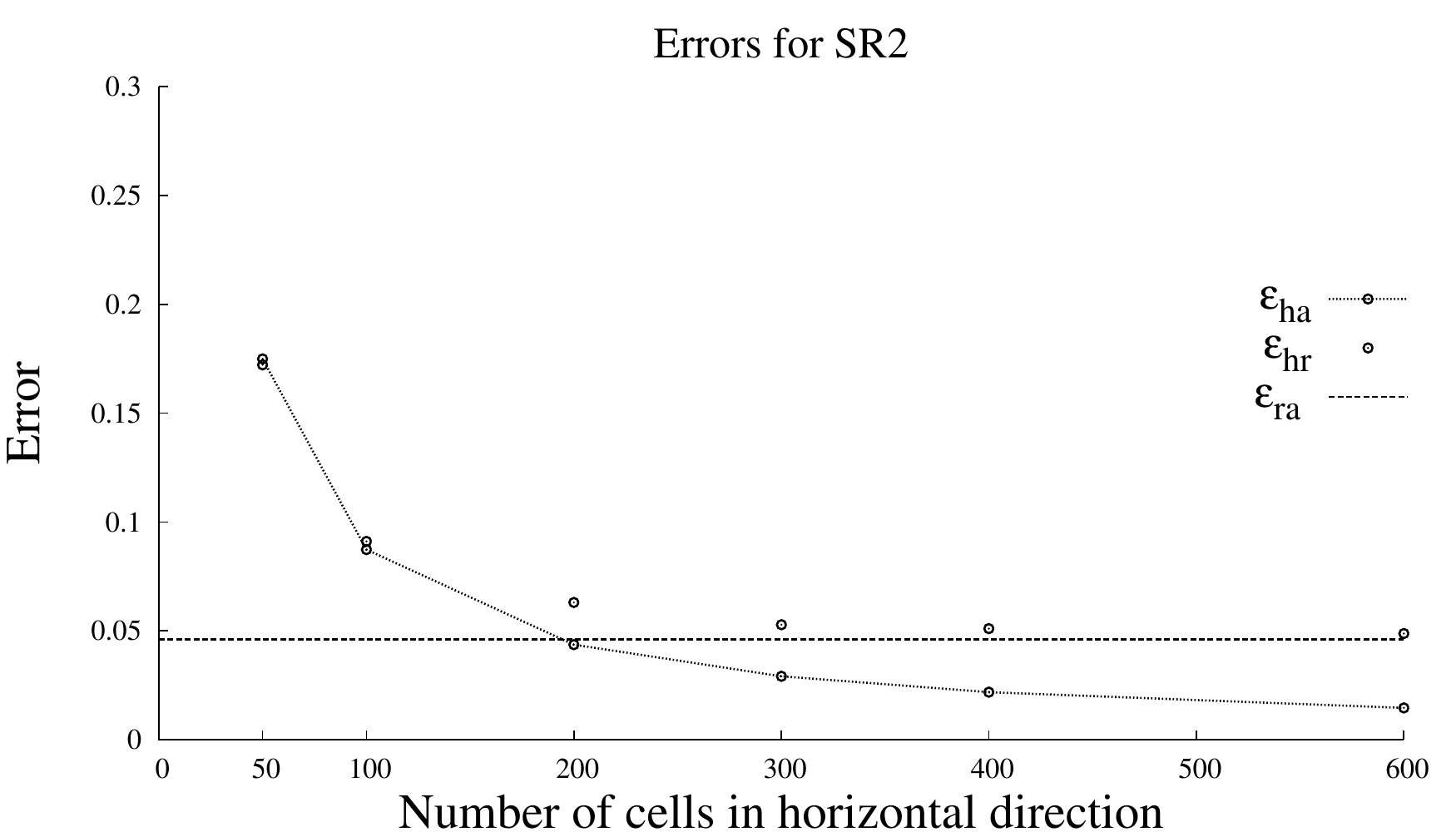}
  \caption{The behavior of the errors $\varepsilon_{hr}$ and
    $\varepsilon_{ha}$ as function of the hexagonal raster
    dimension.}
  \label{errors_fig.01}
\end{figure}

Now, given the raster data $g^r$, one can build a hexagonal
raster following the method described in the previous
section.  To emphasize the approximation behavior of our
method, we construct several hexagonal rasters, each one
being defined by the number of the cells considered along
the $Ox$ direction.  The number of cell rows along the $Oy$
direction results from the requirement to cover the domain
$\mathcal{D}$.  We choose five cases corresponding to $50,
100, 200, 300$ and $600$ horizontal cells.  As an error
measure, we consider the $\mathbb{L}^1$-distance between two
functions. Thus, define
\begin{equation*}
  \varepsilon_{hr} := \frac{1}{\gamma}\int_{\Omega} 
  \left| g^r-g^h \right| {\rm d}\xi, \quad
    \varepsilon_{ha} := \frac{1}{\gamma}\int_{\Omega}
    \left| G-g^h \right| {\rm d}\xi, \quad
    \varepsilon_{ra} := \frac{1}{\gamma}\int_{\Omega}
    \left| G-g^r \right| {\rm d}\xi,
  \label{err}
\end{equation*}
where $\Omega:=\mathcal{D}\cap\widetilde{\mathcal{D}}$,
and $\gamma:=\int_{\Omega}|g^r| {\rm d}\xi$.  The numerical
results for these errors are graphically illustrated in
Figure~\ref{errors_fig.01}, for parameter $a=1$ in $G$.

One should remark that, in general:

a) for a given GIS raster function $g^r$, the error
$\varepsilon_{hr}$ does not go to $0$ as the hexagonal cell
radius goes to $0$;

b) $g^h$ tends to approximate the mathematical function $G$
rather than the approximation $g^r$ of $G$,
$\varepsilon_{hr}>\varepsilon_{ha}$;

c) if the number of cells is high enough, then $g^h$ is a
better approximation of the mathematical model expressed by
function $G$ than its first approximation $g^r$;

d) a better approximation of $G$ by $g^r$ improves the
quality of the entire approximation scheme.

This behavior lies in the fact that $g^h$ approximates
$\widetilde{g}$, while $\widetilde{g}$ tries to recover the
function $G$ (and this holds if $G$ satisfies some
regularity properties).

\begin{figure}[h!]
\centering
\begin{tabular}{cc}\hline
\includegraphics[width=0.45\linewidth]{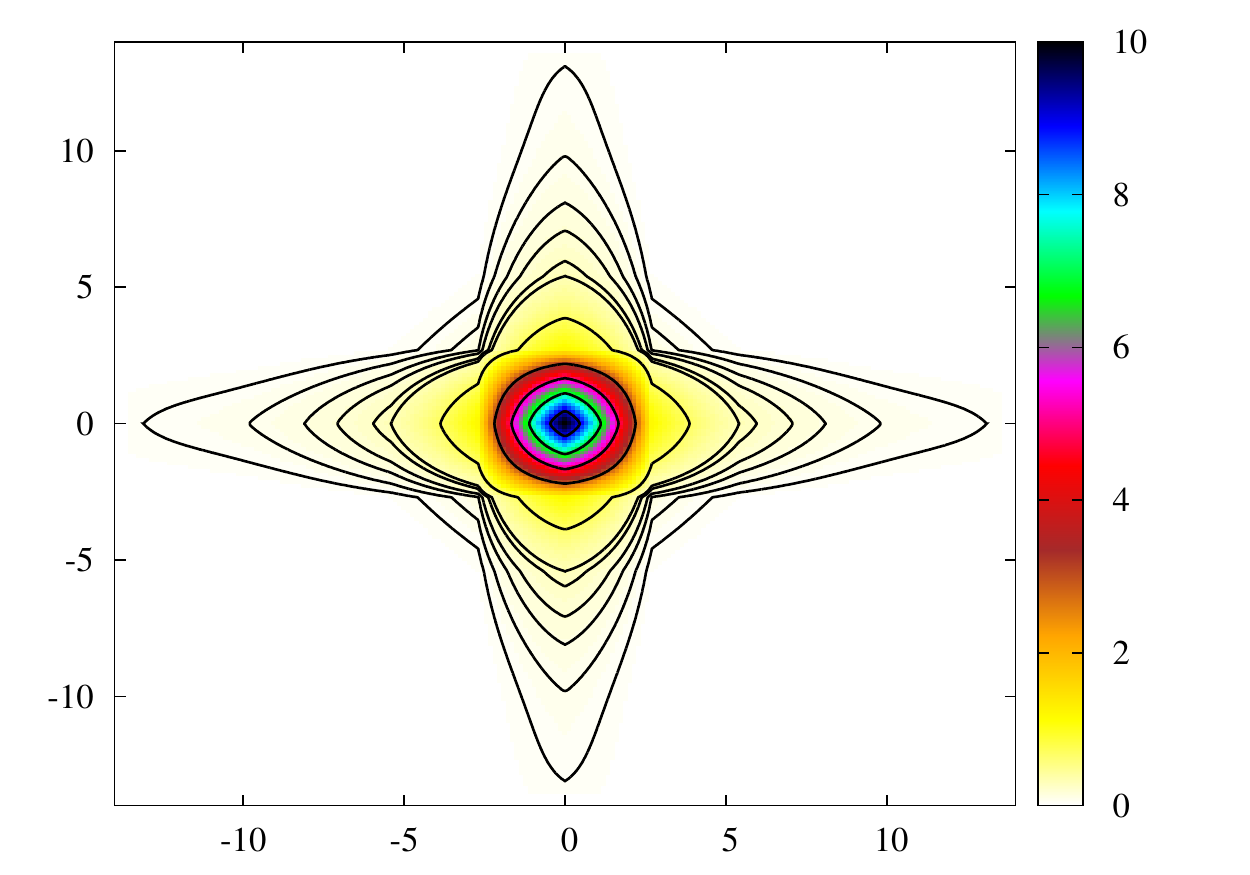}&
\includegraphics[width=0.45\linewidth]{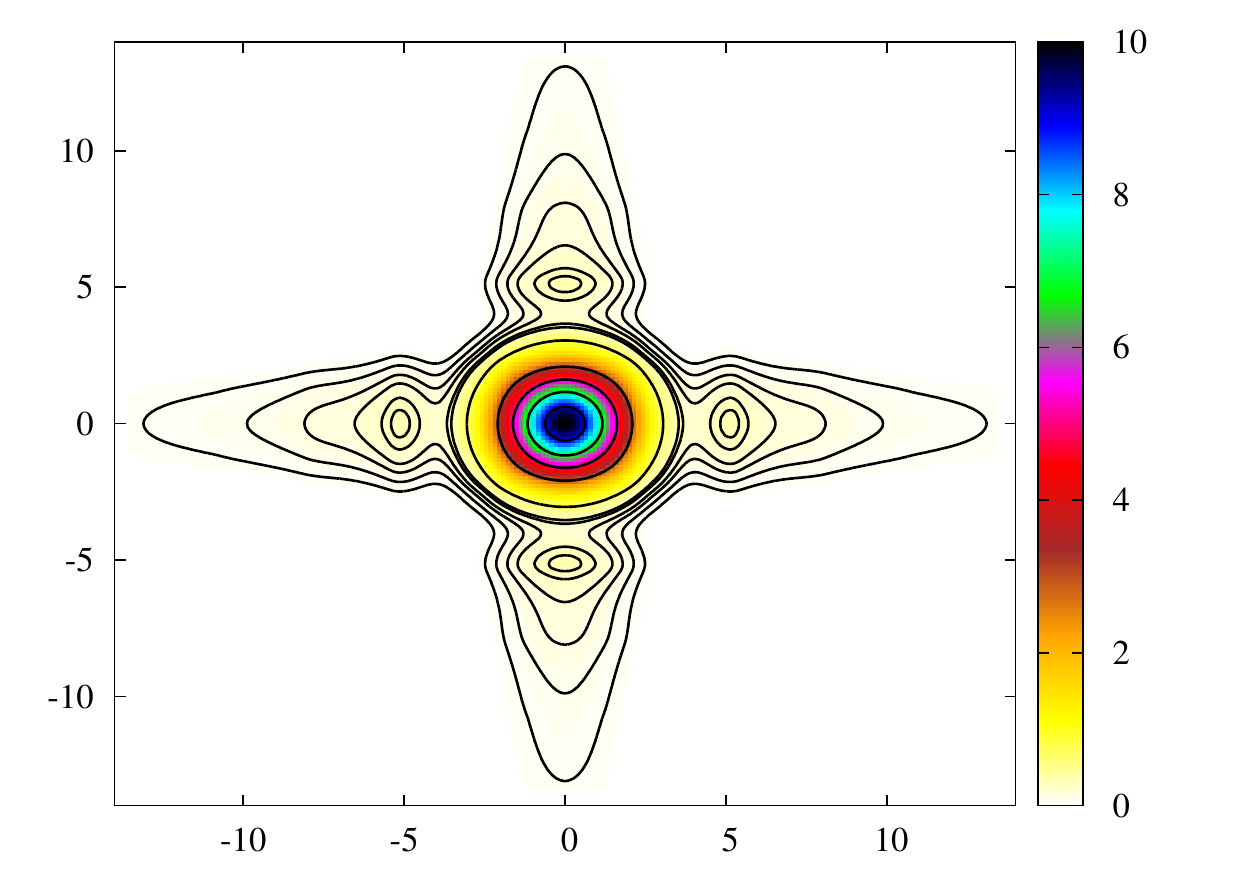}\\
\includegraphics[width=0.45\linewidth]{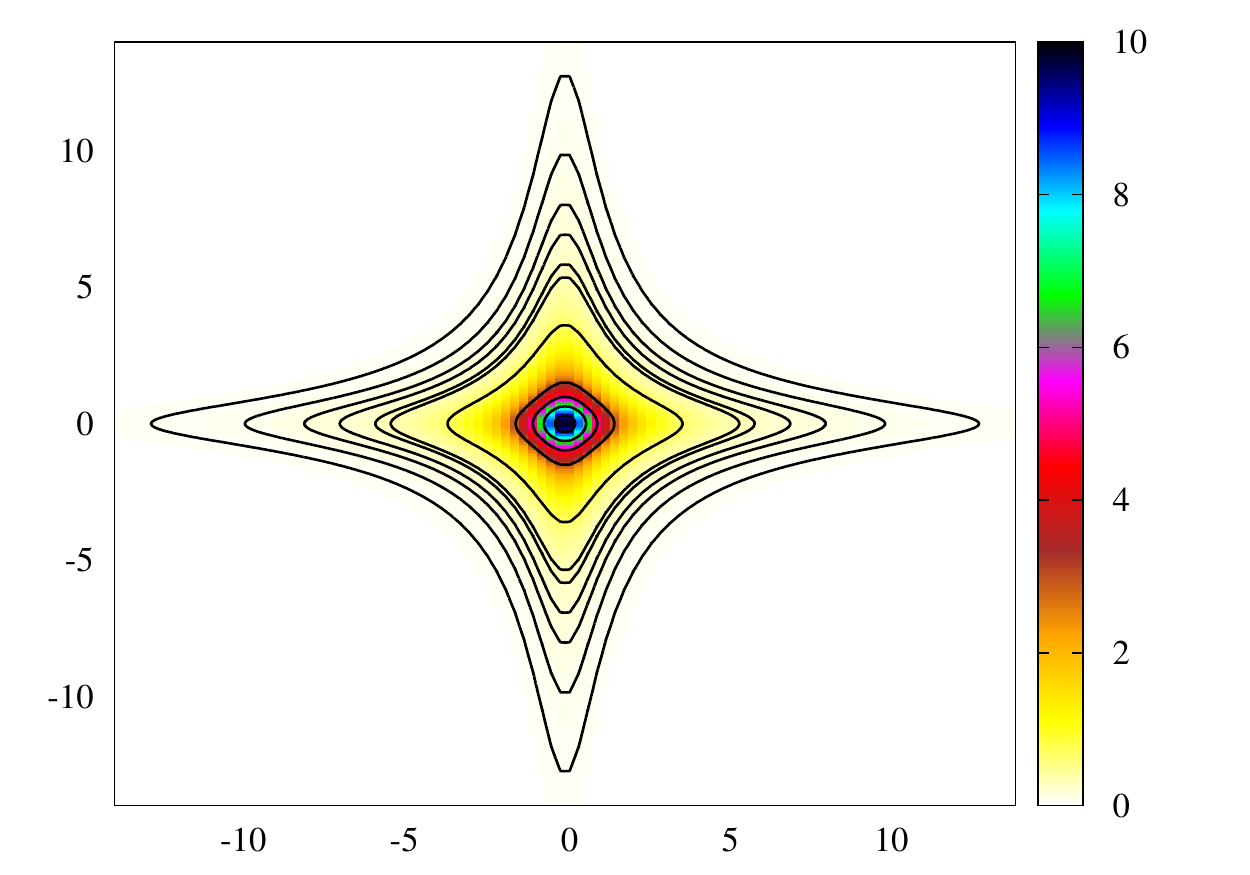}&
\multirow{-7}{*}{
\begin{tabular}{ccc}
  \hline
  &$\varepsilon_{ea}$&$\varepsilon_{er}$\\
  \hline
  {\bf ENO}&$0.447$&$0.5965$\\
  {\bf CRS}&$0.428$&$0.5675$\\
  \hline
\end{tabular}
}
\end{tabular}
\caption{Comparison between 2D ENO and CRS.  The level
  curves of these two extension functions are drawn in the
  upper left and right pictures for ENO and CRS,
  respectively.  The level curves of $G$ are represented in
  the lower left image.  The relative $\mathbb{L}^1$ errors
  of the extension functions, defined as
  $\varepsilon_{ea}=||\widetilde{g}-G(x,y;10)|| / ||g^r||$,
  $\varepsilon_{er}=||\widetilde{g}-g^r|| / ||g^r||$ are
  given in the table.}
\label{fig_ENO_filtering}
\end{figure}

In the next theoretical example, we analyze the fidelity of
representing $G$ with $\widetilde{g}$.  By fidelity, we mean
a characteristic of the method of not introducing spurious
artifacts (bumps, pits, oscillations etc.) which are not
supported by the raster data.  Here, we compare ENO to
filter reconstruction method (Catmull-Rom cubic spline
(CRS), \cite{dodgson}), well known and often used due to
their efficiency in constructing smooth functions from
discrete data.  We mention that both methods use cubic
polynomials and have the same accuracy.  Using a
$10\times10$ raster covering the rectangular domain
$[-14,14]\times[-14,14]$ generated with the Runge type
function $G(x,y;10)$ from (\ref{eq_G_function}), we try to
reconstruct the original function by ENO and CRS methods.
Figure~\ref{fig_ENO_filtering} presents a comparison for
such a case between these two methods.  We have specifically
chosen such a low resolution raster in order to highlight
that a scarcity of the data (as in some environmental
applications) may lead to poor results when the chosen
extrapolation method is inappropriate.  Although the CRS
reconstructed $\widetilde{g}_{CRS}$ is smoother (also the
$\mathbb{L}^1$ error is a slightly better) than the one of
ENO, we remark that $\widetilde{g}_{CRS}$ exhibits
unpleasant artifacts.  We mention that, if a higher
resolution raster is chosen, then both methods give very
good results.

%----------------------------------------------------------
%==========================================================

%==========================================================
\subsubsection{Case study with GIS data}
\label{Section_densit_behav}
In this section, we use GIS data from Ampoi's Valley with
large cells (100 m cell size) and from Paul's Valley - a
subdomain in Ampoi's basin - with much smaller cells (10 m
cell size).

First, we analyze the behavior of our DP
  method with respect to the input data density, using the
  GIS from Paul's Valley.  This represents the basis raster
  we refer in the next example.  One associates a
  reticulated function $g$ to this raster as in
  Section~\ref{Prnd_section}, i.e.
\begin{equation*}
  g:\mathcal{N}\to \mathbb{R}, \; \mathcal{N}=\bigcup_{i,j}
  \left\{ (x_i,y_j) \right\}, \; g(x_i,y_j)=g^r_{i,j},
\end{equation*}
where $(x_i,y_j)$ and $g^r_{i,j}$ are the centers and GIS
values on the squares, respectively.  Starting from this
basis raster, we construct several rasters by randomly
eliminating part of the data (replacing them with
NODATA\_VALUE), this process being achieved with some
restrictions described in \ref{ApB}.  Then we associate a
reticulated function $g_{(\alpha)}$ to such a raster
$\alpha$,
\begin{equation*}
  g_{(\alpha)}:\mathcal{N}_{(\alpha)}\to \mathbb{R}, \;
  \mathcal{N}_{(\alpha)}\subset\mathcal{N}, \; g_{(\alpha)}
  = g \textnormal{ on } \mathcal{N}_{(\alpha)},
\end{equation*}
where $\mathcal{N}_{(\alpha)}$ is the set of the remaining
square centers.

To test the capability of ENO method to recover the missing
data, we apply it to $g_{(\alpha)}$ for constructing the
extension function $\widetilde{g}_{(\alpha)}$ by means of
which we can calculate the values at the points from
$\mathcal{N}\setminus\mathcal{N}_{(\alpha)}$.  Now, one has
to compare these values with the corresponding ones from the
basis raster.  For this purpose, we use two different
measures:
\begin{enumerate}
\item the root-mean-square error\footnote{Since ENO is an
    exact interpolation method, the expressions $g^r_{i,j} -
    \widetilde{g}_{(\alpha)}(x_i,y_j)$ obviously vanish for
    $(x_i,y_j)\in\mathcal{N}_{(\alpha)}$.} (RMSE)
    \begin{equation*}
      RMSE =
      \sqrt{\frac{1}{\nu}\sum_{i,j} \left(
          g^r_{i,j} - \widetilde{g}_{(\alpha)}(x_i,y_j) \right)^2},
    \end{equation*}
    where\footnote{The notation $\#A$ for some set $A$
      stands for the cardinal of $A$.}
    $\nu=\#\left(\mathcal{N}\setminus\mathcal{N}_{(\alpha)}\right)$;
  \item the infinity norm of $g - \widetilde{g}_{(\alpha)}$ on $\mathcal{N}$
    \begin{equation*}
      ||g-\widetilde{g}_{(\alpha)} ||_{\infty} = \max\limits_{i,j}
        | g^r_{i,j} - \widetilde{g}_{(\alpha)}(x_i,y_j)| .
    \end{equation*}
\end{enumerate}

Then, the accuracy of DP is analyzed by comparing the basis
hexagonal raster $g^h$ generated by $\widetilde{g}$ with the
hexagonal rasters $g^h_{(\alpha)}$ generated by
$\widetilde{g}_{(\alpha)}$.  These measures are calculated
for different cases $\alpha$ and presented in
Table~\ref{tab_err}.
\begin{table}[htbp!]
  \centering
  \caption{This table presents the errors
    between a basis square raster and different
    test square rasters $(\alpha)$ by means of RMSE
    and $\| \cdot \|_{\infty}$ in columns 3 and 4,
    respectively.  The errors between the basis hexagonal
    raster (built from the basis square raster using ENO)
    and the test hexagonal rasters (built from test square
    rasters using again ENO) are given in columns 5 and 6.  
    The basis square raster with cell size of 10 m contains 
    $361\times 561$ GIS data from Romanian Paul's Valley, 
    while the hexagonal rasters are based on cells of 
    10.418 m size.}
    \begin{tabular}{cc cccc}
      \cline{1-6} & & \multicolumn{2}{c}{Square Rasters} &
      \multicolumn{2}{c}{Hexagonal Rasters} \\
      \multicolumn{1}{c}{$\alpha$} & $\#\mathcal{N}_{(\alpha)} / \#\mathcal{N}$ & RMSE & $||\cdot ||_{\infty}$ & RMSE &
      $||\cdot ||_{\infty}$\\
      \hline
      $1$&$0.25$&$0.097$&$2.33$&$0.076$&$2.230$\\
      $2$&$0.21$&$0.106$&$2.50$&$0.086$&$2.500$\\
      $3$&$0.17$&$0.121$&$3.40$&$0.102$&$3.420$\\
      $4$&$0.14$&$0.123$&$2.66$&$0.109$&$1.957$\\
      $5$&$0.12$&$0.137$&$3.73$&$0.122$&$3.370$\\
      \hline
    \end{tabular}
  \label{tab_err}
\end{table}

As we have commented in the beginning of
Section~\ref{sect_numerical_applications}, it is almost
impossible to measure the error between the raster $g^h$ and
the physical quantity $G$ modeled by it (we do not know this
function).  However, we can perform an analysis of the
accuracy when we have two different rasters of the same
area, one at a low and the other at high resolutions.  It is
assumed that the high resolution raster is a good
approximation of $G$.  In such case, we generate the
extension function from the low resolution raster, and then
we compare it to the high resolution raster function.
Following this strategy, we therefore perform a qualitative
and then a quantitative analysis.

\begin{figure}[htbp!]
  \centering
  \includegraphics[width=0.3\linewidth]{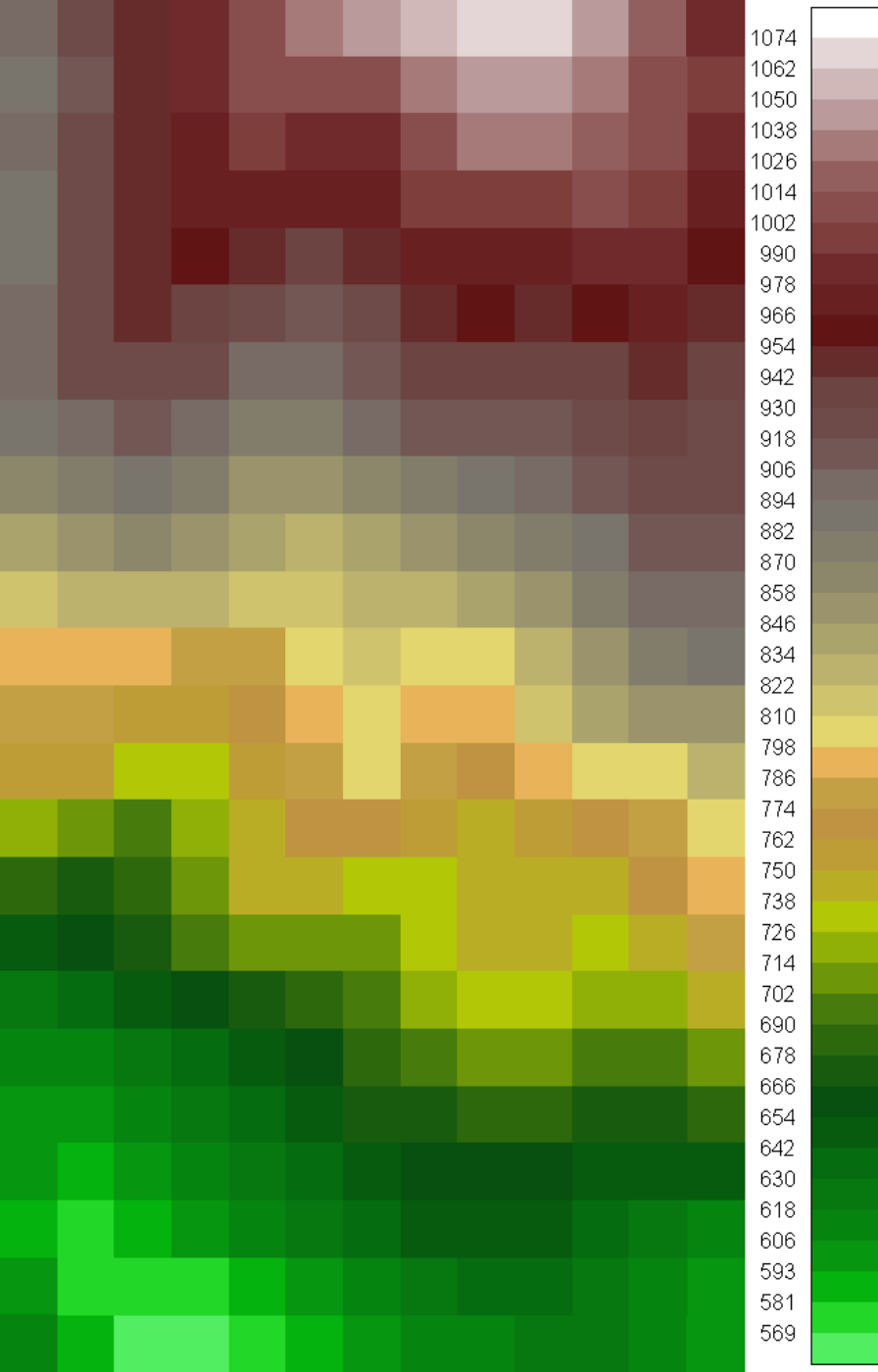}
  \hspace{3mm}
  \includegraphics[width=0.3\linewidth]{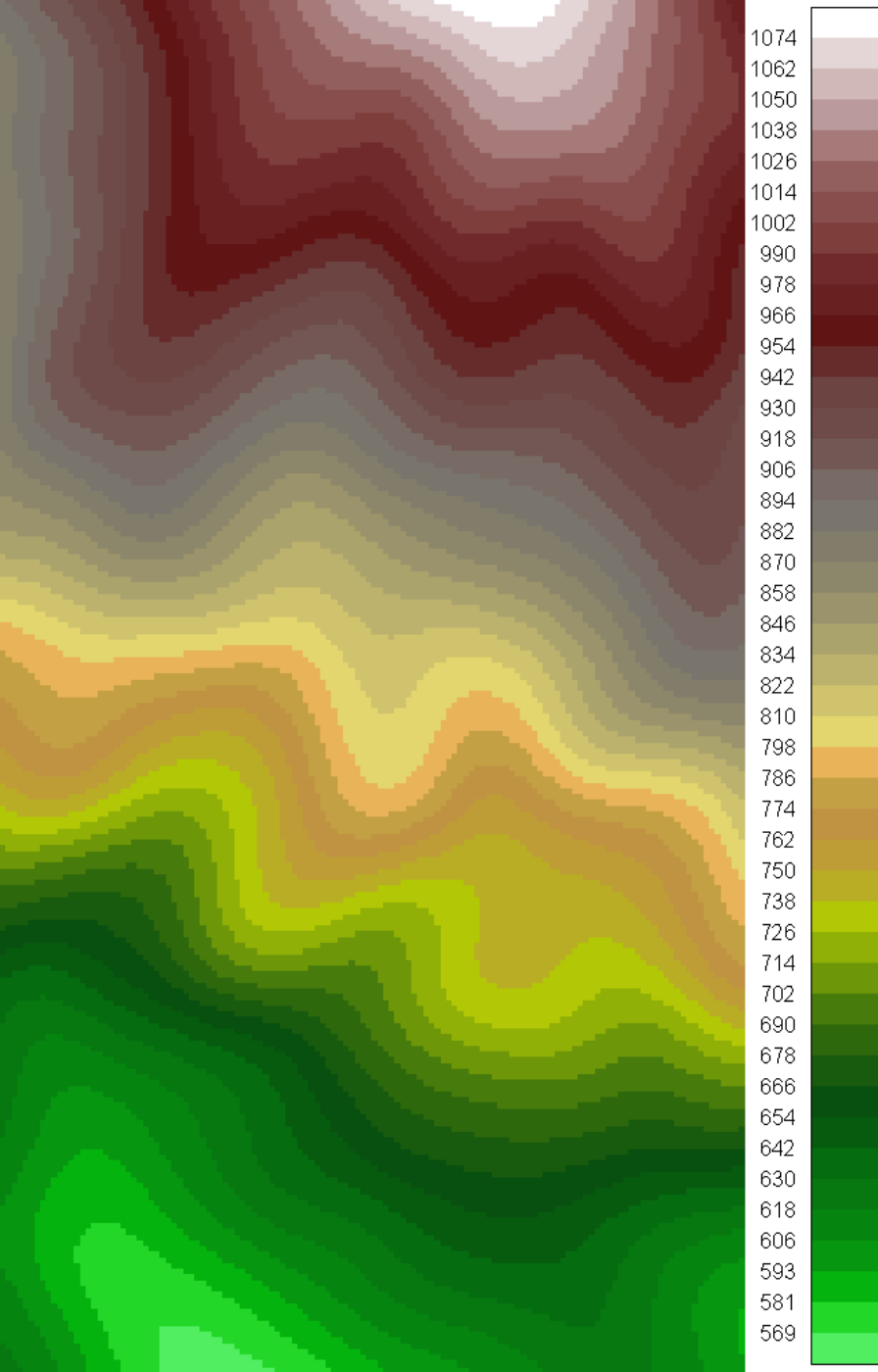}
  \hspace{3mm}
  \includegraphics[width=0.3\linewidth]{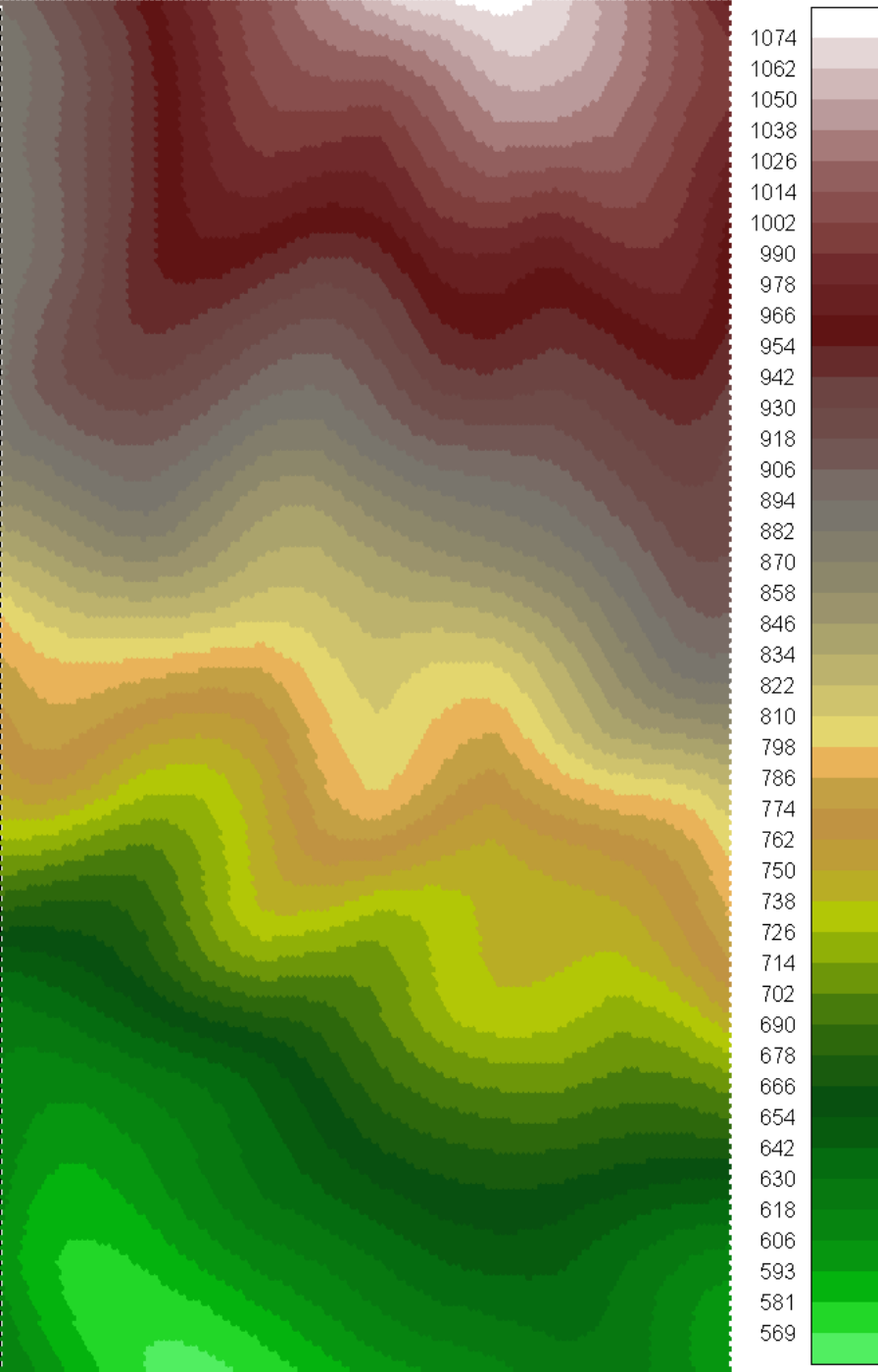}
  \caption{Relief from Romanian Paul's Valley.  All the
    figures represents the same zone from Romanian Paul's
    Valley and are obtained as follows: the left one from a
    $13\times 24$ GIS raster data with square cells of 100 m
    size, the middle one from a $130\times 240$ GIS raster
    data with square cells of 10 m size.  The figure on the
    right is obtained on a hexagonal raster (of 5.7735 m
    cell size) applying our ENO method to the same data
    input as for the first figure.}
  \label{fig_Paul_relief}
\end{figure}

Figure~\ref{fig_Paul_relief} presents pictures of the same
portion of Paul's Valley terrain obtained from different
raster data.  In this figure, as well as in Figures
\ref{fig_ampoi} and \ref{fig_paul_water_routing}, the color
codes correspond to altitudes which are measured in meters.
The left and the middle images are generated by rectangular
rasters of 100 and 10 m cell size, respectively.  The right
image ``pictures'' the accuracy of our DP method with
respect to the ``real'' function $G$ behind the data rasters
and it is build on a hexagonal raster (of 5.7735 m cell
size) obtained from the low resolution rectangular raster
corresponding to the left image.  The resemblance of this
figure to the one in the middle which is built from a much
denser GIS raster represents an argument that our DP is able
to recover the ``real'' function $G$.  This result is
similar to the theoretical cases previously studied.
Moreover, this resemblance can be also verified with the
airplane photo from Figure~\ref{fig_paul_water_routing}.

The quantitative results of our analysis are summarized in
Table~\ref{table_errors_ampoi_paul}.
\begin{table}[!htbp]
  \centering
  \caption{Error analysis for the case study with GIS data.
    This table contains the relative errors of the extension
    functions provided by ENO, CRS, and Id (the extension
    function of Id is identical to the raster function).
    The analysis is performed using the same two GIS rasters
    (one of high (H) and the other one of low (L) resolution)
    as those in Figure~\ref{fig_Paul_relief}.\\
    $\varepsilon^{L}_{L}=||\widetilde{g}_{L}-g^{r}_{L}|| / ||g^r_{L}||$,
    $\; \varepsilon^{L}_{H}=||\widetilde{g}_{L}-g^{r}_{H}|| / ||g^r_{H}||$,
    $\; \varepsilon^{H}_{H}=||\widetilde{g}_{H}-g^{r}_{H}|| / ||g^r_{H}||$,
    where the norms $||\cdot||$ are in $\mathbb{L}^1$.
    The extension functions $\widetilde{g}_{L}$,
    $\widetilde{g}_{H}$ are generated by the low, and high
    resolution raster functions $g^{r}_{L}$ and $g^{r}_{H}$,
    respectively.}
  \begin{tabular}{cccc}
    \hline
    &$\varepsilon^{L}_{L}$&$\varepsilon^{L}_{H}$&$\varepsilon^{H}_{H}$\\
    \hline
    {\bf ENO}&$0.0076$&$0.0033$&$0.00078$\\
    {\bf CRS}&$0.0075$&$0.0033$&$0.00078$\\
    {\bf Id} &$0.0000$&$0.0081$&$0.00000$\\
    \hline
  \end{tabular}
  \label{table_errors_ampoi_paul}
\end{table}
One should remark that:

a) for ENO and CRS, the extension function is closer to the
high than lower resolution raster function, although the
extension function is build from the lower resolution raster
function: $\varepsilon^{L}_{H} < \varepsilon^{L}_{L}$.  This
explains why the right picture from
Figure~\ref{fig_Paul_relief} is more similar to the middle
picture than to the left one.

b) the error between the extension and the high resolution
raster function is smaller than the error between the high
and the low resolution raster functions:
$\varepsilon^{L}_{H}(ENO), \varepsilon^{L}_{H}(CRS) <
\varepsilon^{L}_{H}(Id)$.  This explains why the middle
picture from Figure~\ref{fig_Paul_relief} is more similar to
the right picture than to the left one.

c) the extension function generated by the high resolution
raster is closer than the extension function generated by
the low resolution raster to the high resolution raster
function: $\varepsilon^{H}_{H} < \varepsilon^{L}_{L}$.  This
says that the extension function generated by the high
resolution raster is better than the extension function
generated by the low resolution raster.
%==========================================================

%==========================================================
\subsection{Water Routing}
\label{Section_WaterRouting}
By water routing one means a method to define the way cells
exchange water between one another or the way water flows
along the entire cellular.  There exists a large body of
literature devoted to the subject, see for example
\cite{callagan, moore, quinn, soille, tarboton, wilson}, but
most of the existent schemes are designed for the square
cells.  However, there are also a few papers (see
\cite{ambrosio01} for example) where hexagonal instead of
square cellulars are used and different water routing
schemes are presented.  Any water routing scheme assumes two
kinds of rules: qualitative and quantitative ones.  The
qualitative rules give us a picture of the water path, while
the quantitative rules give us information about the amount
of water discharge along the water path.  In what follows,
using notions as {\it donor cells}, {\it receptor cells},
{\it steepest descent}, and {\it water velocity of a cell},
we present a method to model the water path.

\begin{figure}[htbp!]
  \centering
  \includegraphics[width=0.3\linewidth]{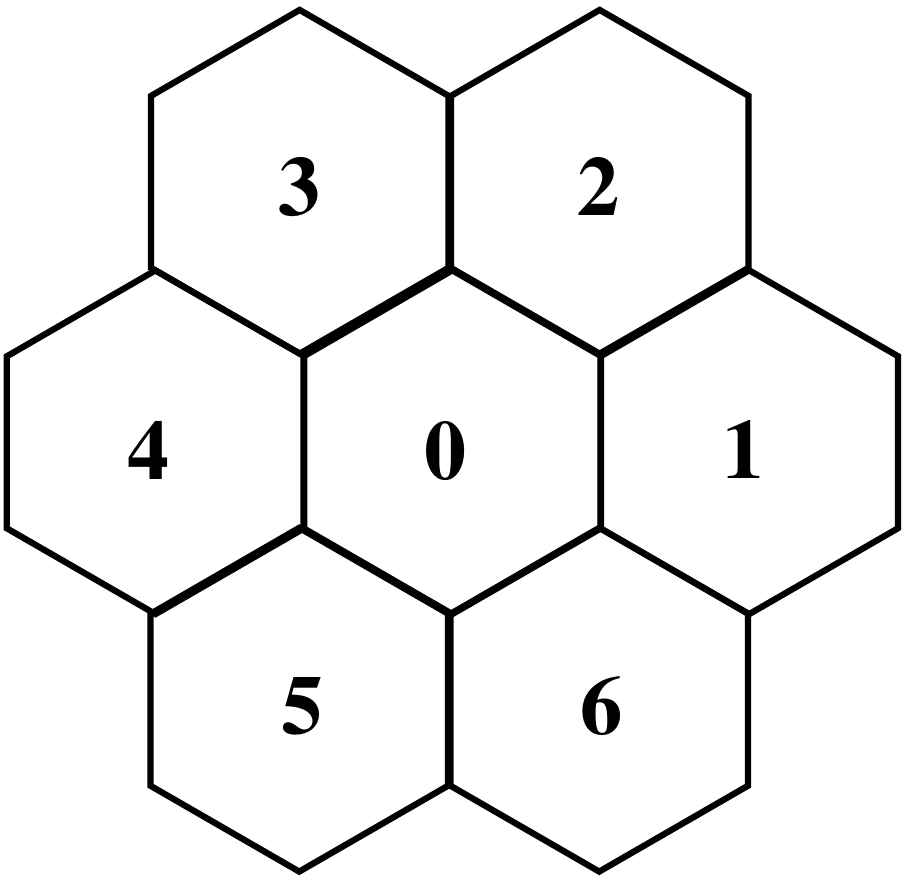}
  \caption{A central cell $i$ (indexed by $0$) and its
    neighbors (indexed from $1$ to $6$).}
  \label{rosetta_fig}
\end{figure}

First, let us introduce the local configuration of a
hexagonal cellular.  By local configuration one means a
central cell with its six adjacent cells defining the
neighborhood of this central cell.  The sides of the central
cell are indexed from 1 to 6 counterclockwise.  One denotes
by ${\mathcal V}_i$ the neighbors of a cell $i$.  The
quantities referring to the central cell are indexed with
$0$ and those referring to its neighbors are indexed from 1
to 6, see Figure~\ref{rosetta_fig}.

%\begin{minipage}{\linewidth}
%\end{minipage}

{\it Steepest descent and water velocity}.  For each cell
$i$, define its water potential $\psi_i$ to be
\begin{equation*}
  \psi_i=h_i+z_i,
\end{equation*}
where $h_i$ denotes the water depth in this cell and $z_i$
stands for the altitude to its soil surface.

We now consider a reference cell $i$ of center $(x_0,y_0)$
and its local configuration with the cells of centers
$\{(x_j,y_j)\}_{j=\overline{1,6}}$.  One may approximate the
tangent plane $\Pi$ to the surface of the water potential by
the best interpolant plane given by the points
$\{(x_j,y_j,\psi_j)\}_{j=\overline{0,6}}$.  The equation of
this plane can be described as
\begin{equation*}
  z-z_0=a(x-x_0)+b(y-y_0),
\end{equation*}
where
\begin{equation*}
  \label{rules_flow_1.eq}
  \begin{array}{l}
    a=\displaystyle\frac{1}{6\sqrt{3}r},
    \left(2(\psi_1-\psi_4)+\psi_2-\psi_5+\psi_6-\psi_3\right),
    \\
    b=-\displaystyle\frac{1}{6r}\left(\psi_2-\psi_5+\psi_3-\psi_6\right)
  \end{array}
\end{equation*}
and $r$ represents the radius of the circumscribed circle of
the hexagon.

The steepest descent of the cell $i$ is defined as the
projection of the gravitational force onto its tangent plane
$\Pi$, and therefore, as a vector in $\mathbb{R}^3$ it is
given by
\begin{equation*}
  \boldsymbol{d}=\left(-a,-b,-(a^2+b^2)\right)^{\rm T}.
\end{equation*}

The slope of the cell $i$ is now given by
\begin{equation}
  \label{rules_flow_3.eq}
  s=\sqrt{\displaystyle\frac{a^2+b^2}{1+a^2+b^2}},
\end{equation}
and therefore, the projection versor of the steepest descent
onto the $xOy$ plane is given by
\begin{equation}
  \label{rules_flow_2.eq}
  \boldsymbol{\tau}=\left(-\displaystyle\frac{a}{\sqrt{a^2+b^2}}, 
    -\displaystyle\frac{b}{\sqrt{a^2+b^2}}\right).
\end{equation}  

{\it Receptor/Donor cells}.  A cell of index $j$ in the
neighborhood ${\mathcal V}_i$ of the central cell $i$ is
called a {\it receptor cell} if it satisfies two conditions:
\begin{equation}
  \begin{array}{l}
    \psi_j<\psi_0\\
    \boldsymbol{\tau}\cdot\boldsymbol{n}_{ij}>0
  \end{array}
\end{equation}
where $\boldsymbol{n}_{ij}$ is the outward unitary normal to
the face $j$ of central cell $i$.  The set of all these
receptor cells $j$ associated to the cell $i$ will be
denoted by $J^{-}_i$. If $J^{-}_i\neq\emptyset$, then $i$ is
called a {\it donor cell}.

{\it Water velocity.}  The water velocity is calculated
using the descent direction (\ref{rules_flow_2.eq}), the
slope (\ref{rules_flow_3.eq}) and a Manning type empirical
law
\begin{equation}
  \label{rules_flow_4.eq}
  \boldsymbol{w}_i=v(h_i,s_i)\boldsymbol{\tau}_i,\,\, v(s,h)=\displaystyle\frac{h^{2/3} s^{1/2}}{n_{M}},
\end{equation}   
where $n_{M}$ is the Manning coefficient and $s$ the slope
of the central cell.

\begin{figure}[hbpt!]
  \centering
  \includegraphics[width=0.4\linewidth]{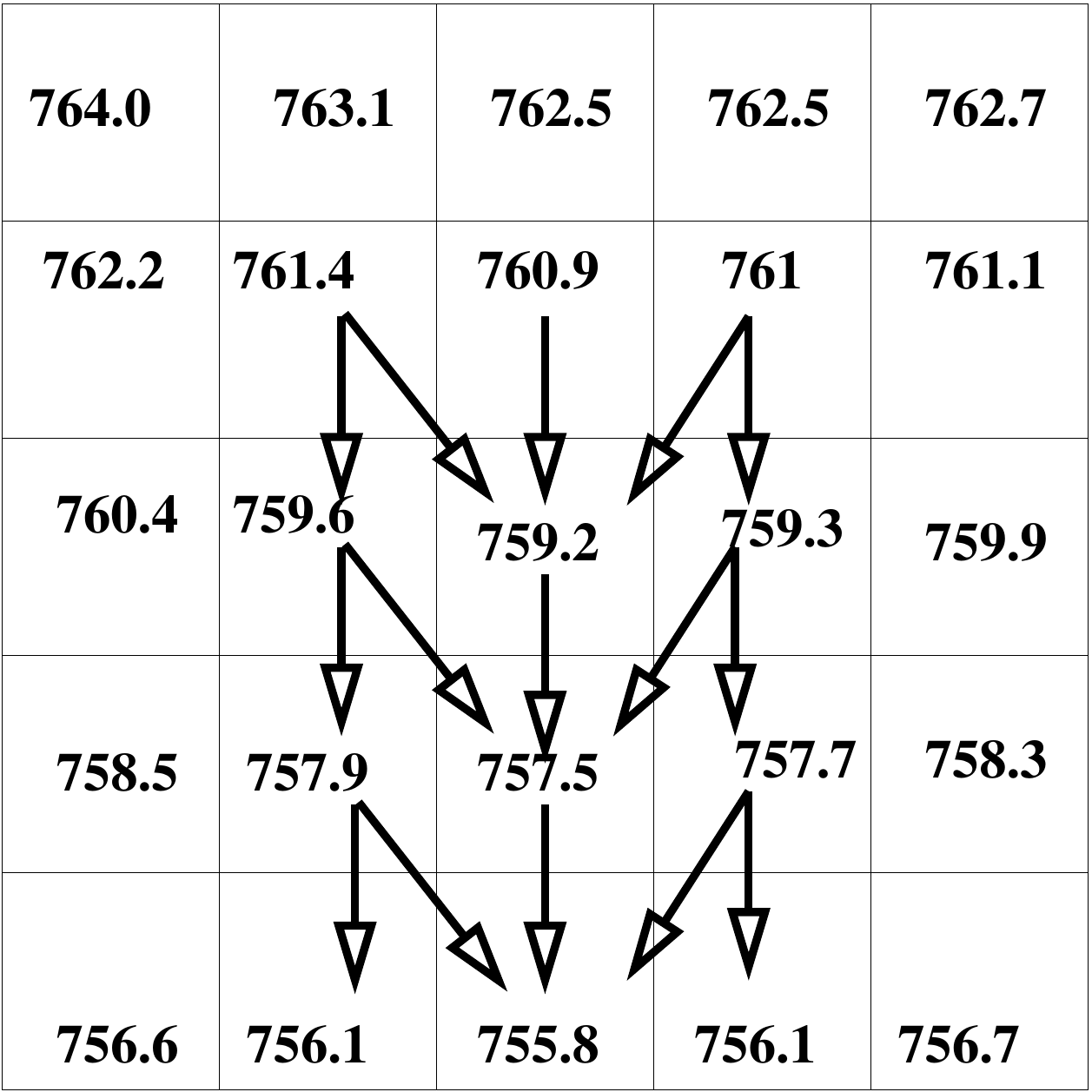}
  \hspace{5mm}
  \includegraphics[width=0.4\linewidth]{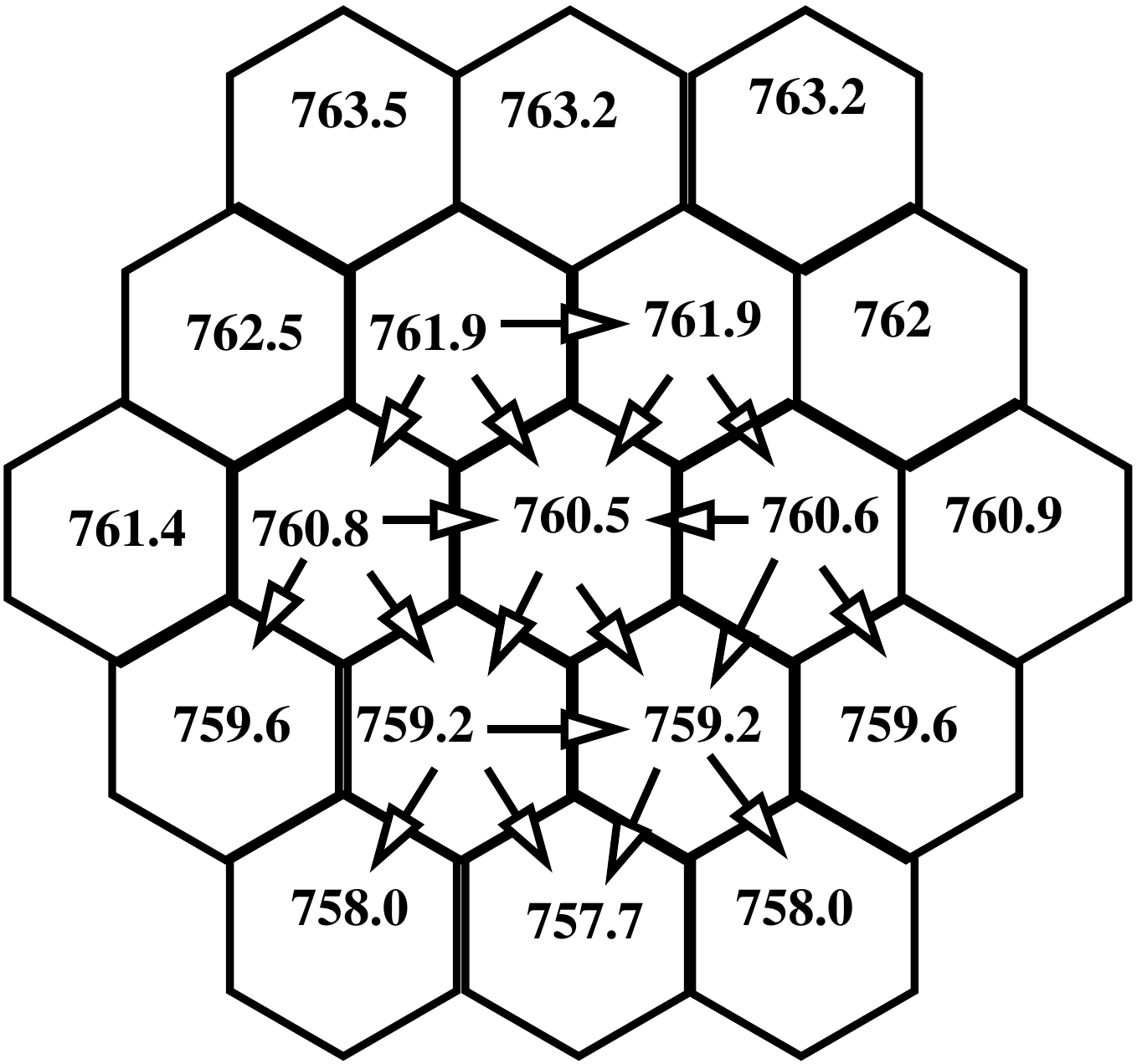}
  \caption{Water flow among cells in rectangular (left side)
    and hexagonal (right side) rasters using Tarboton and
    our rules, respectively.  Data written inside cells
    represent altitudes and they come from a small region
    (the same on the left and on the right) of Romanian Paul
    Valley. The values of the rectangular raster slightly
    differ from those of the hexagonal one because they
    represent two different models of the same terrain
    surface.}
  \label{fig_water_change_comparison}
\end{figure}

{\it Water flux.} The basic assumption on the water flux
across the cells is that the water of any cell flows out to
the neighboring cells toward which the water velocity of
that cell points to, and the water enters a cell only from
the neighboring cells whose water velocities point to that
cell.  This assumption is analogous to the upwind scheme
used in hyperbolic system approximation theory
\cite{gallouet}.  A comparative portrait between Tarboton
method on rectangular raster and the above presented method
for water flow among cells is illustrated in
Figure~\ref{fig_water_change_comparison}.

Let $i$ be a donor cell.  The water lost by the cell $i$
during the time interval $\triangle t$ is given by
\begin{equation}
  \label{rules_flow_5.eq}
  \triangle t\, l\, h_i \sum\limits_{j\in J^{-}_i}\boldsymbol{n}_{ij}\cdot\boldsymbol{w}_i,
\end{equation}
where $l$ is the side of the hexagonal cell.  A receptor
cell $j$ from the set $J^{-}_i$ receives a water quantity
equal to
\begin{equation}\label{rules_flow_6}
  \triangle t\, l\, h_i\, \boldsymbol{n}_{ij}\cdot\boldsymbol{w}_i.
\end{equation}

Thus, the water flux trough the boundary of a cell $i$
during the time interval $\triangle t$ is given by
\begin{equation}
  \label{the_flux_water_for_cell_i}
  {\cal F}_i^{f} = \triangle t\,l\left(- h_i
    \sum\limits_{j\in
      J^{-}_i}\boldsymbol{n}_{ij}\cdot\boldsymbol{w}_i+
    \sum\limits_{k\in {\mathcal
        V}^{+}_i}h_k\boldsymbol{n}_{ki}\cdot\boldsymbol{w}_k\right),
\end{equation}
where ${\mathcal V}^{+}_i$ is the set\footnote{${\mathcal
    V}^{+}_i=\left\{k\in{\mathcal V}_i\left| \; \exists j\in
      J^{-}_k\,\, {\rm s.t.}\,\, i=M(j,k)\right.\right\}$,
  where $M(j,k)$ represents the index of that cell from
  ${\mathcal V}_k$ sharing with $k$ the common face $j$.} of
neighboring cells of $i$ shedding water to this cell.

One can now define the water flow as an iterative process by
\begin{equation}
  \label{eq_iterative_process}
  h^{n+1}_i = h^n_i + {\cal F}_i^{f,n},
\end{equation}
where ${\cal F}_i^{f,n}$ represents the water flux trough
the boundary of a cell $i$ during the time interval
$\triangle t$ at the $n^{\textnormal{th}}$ iteration.

\begin{figure}[htbp!]
  \centering
  \includegraphics[width=.99\linewidth]{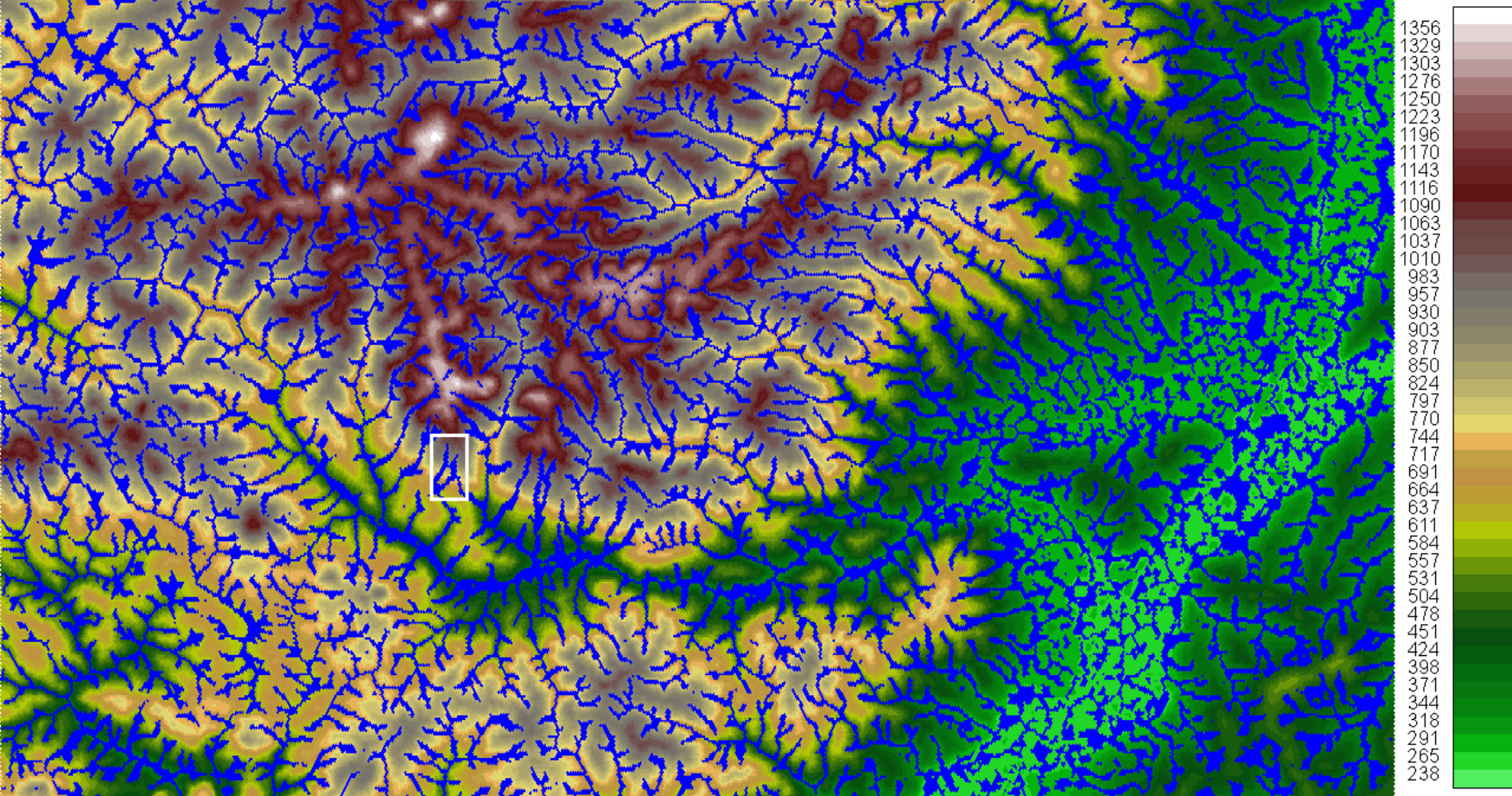}
  \caption{Water accumulation zones in Ampoi's hydrographic
    basin.  Starting from the real elevation GIS raster data
    (542 columns by 310 rows of cells with 100 m side
    length) of Ampoi's landscape, we captured the areas of
    concentrated water flow using a hexagonal cellular
    automaton.  The hexagonal raster (with cell side of
    57.735 m) was obtained by porting the GIS data following
    the method described in Section~\ref{sect_porting_data}.
    Beginning with a uniform shallow water level for the
    entire landscape, we processed for a set time interval
    the water flow across the cells using the law described
    by (\ref{the_flux_water_for_cell_i}).  The blue area in
    the figure represents the hexagonal cells where the
    water layer exceeds the initial level.  The white border
    rectangle inside this figure marks the area from Paul's
    Valley we have referred in Figures \ref{fig_Paul_relief}
    and \ref{fig_paul_water_routing}.}
  \label{fig_ampoi}
\end{figure}

Figure~\ref{fig_ampoi} is a snapshot of a water flow modeled
using the method described in this section.  

\begin{figure}[htbp!]
  \centering
  \includegraphics[height=0.49\linewidth]{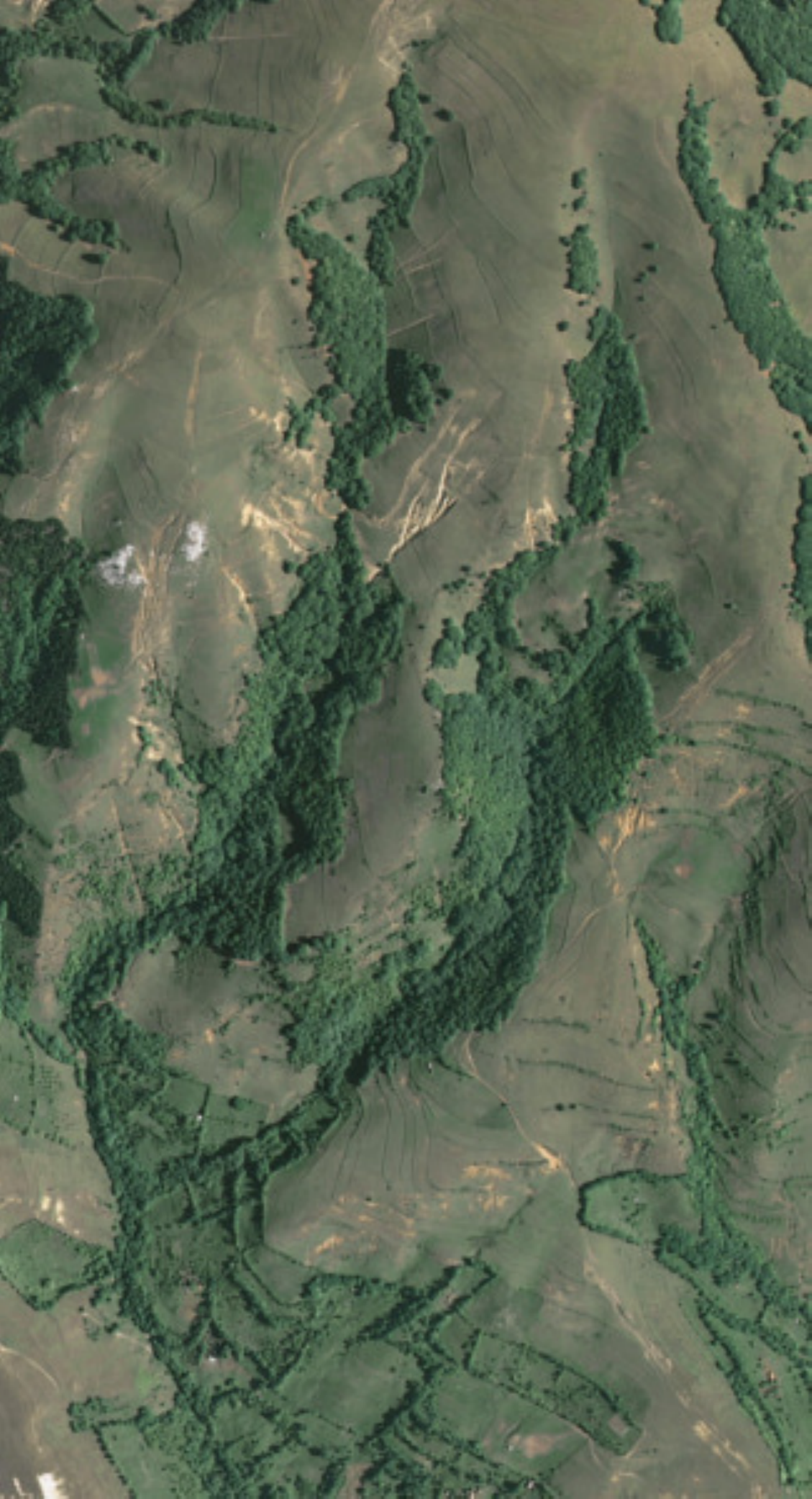}
  \hspace{3mm}
  \includegraphics[height=0.49\linewidth]{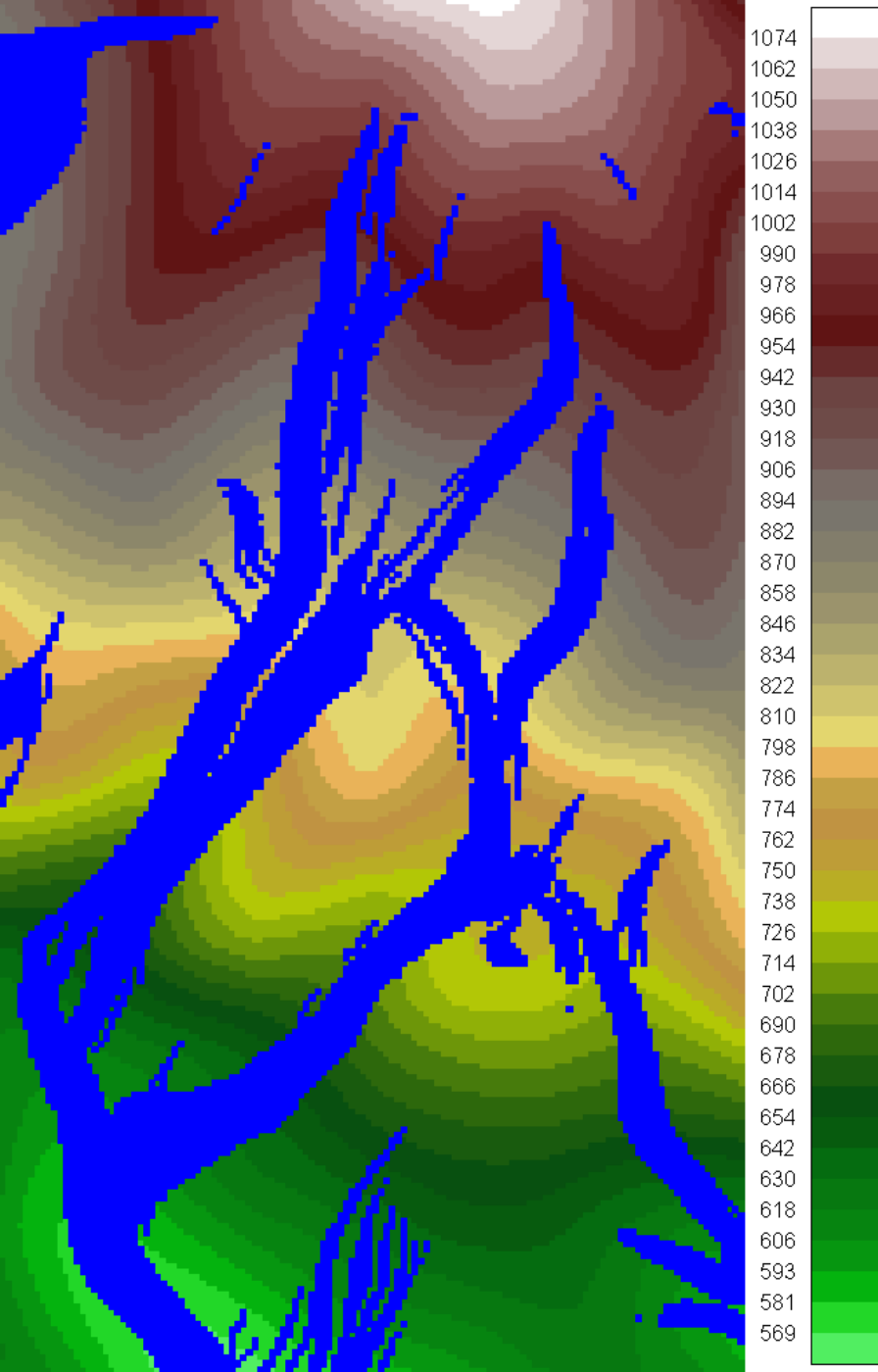}
  \hspace{3mm}
  \includegraphics[height=0.49\linewidth]{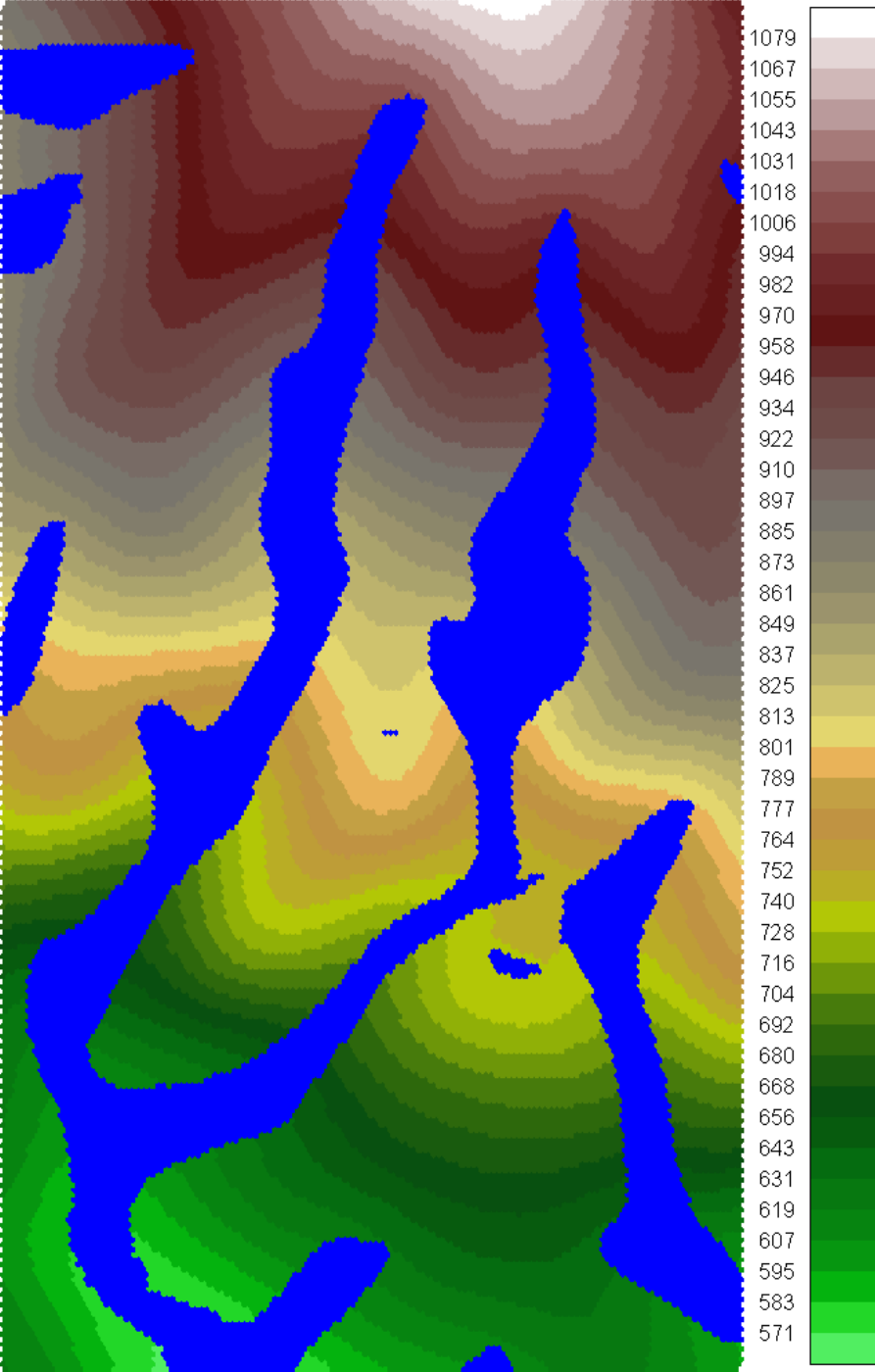}
  \caption{Potential water accumulation zones in Paul's
    Valley.  The left figure is a photo of this region taken
    from the airplane.  The relief from the image in the
    middle is given by the same $130\times 240$ GIS raster
    data with square cells of 10 m size used for the middle
    picture from Figure~\ref{fig_Paul_relief}.  The relief
    from the right picture is generated from the hexagonal
    raster (of 5.7735 m cell size) obtained by applying ENO
    DP to the same GIS raster used for the middle image.
    The blue transparent areas represent the potential water
    accumulation zones determined by an iterative process of
    type (\ref{eq_iterative_process}).  The picture in the
    middle is obtained using Tarboton's rules for water
    change among cells.  For the picture on the right side,
    we used our water routing method described in this
    section.  One clearly observes that the accumulation
    zones indicated by this method overlap almost perfectly
    the ravines and are less spread than the ones identified
    in the middle image.}
  \label{fig_paul_water_routing}
\end{figure}

Figure~\ref{fig_paul_water_routing} contains three images of
the same zone cut out from Paul's Valley.  The first one is
a photo where one can observe the ravines of this region,
while the other two are images constructed from GIS data and
include the water accumulation zones found in two different
ways.  It is known that if Tarboton's rules are used, then
there is the possibility that the flow direction in some
cells cannot be defined.  To overcome this problem, Tarboton
uses special treatments for such cells.  The iterative
process (based on Tarboton) implemented by us only uses the
basic rules without any improvement and this can be a reason
of why the water accumulation zones from the picture in the
middle are so spread out.
%==========================================================

%==========================================================
\section{Conclusions and Remarks}
We provided a method for porting rectangular raster data on
to hexagonal raster.  The basic idea of the algorithm is to
use an intermediate function $\widetilde{g}$ that
interpolates the rectangular raster function $g^r$ and then
define the hexagonal raster function $g^h$ as the point
values of $\widetilde{g}$ at the center of the hexagonal
cell.  The interpolation method can be extended to a less
regular grid and can be also used (see \ref{ApA}) to recover
missing data or to filter outliers.  Using some theoretical
as well as some practical examples, we show that the
extension function is closer to the function represented by
raster data than raster function itself.  As comparing to a
well known cubic spline interpolant (CRS), Catmull-Rom (see
Figure~\ref{fig_ENO_filtering}), ENO and CRS method have the
same degree of accuracy, but CRS is faster than ENO method, 
and furnishes a smooth
function in contrast with almost continuous function
obtained by ENO method (see also Remark~\ref{remark1} in
\ref{ApA} for more theoretical details).  The main argument
for using ENO instead of CRS is that ENO method does not
introduce spurious oscillations for cases of data with large
gradients and works on a less regular grid points.  For the
case of digital terrain model, it is crucial to preserve the
terrain shape and not to introduce artifacts.  Both methods
also provide a good compromise between a reduced amount of
computational effort and the accuracy of the extended
function.

The OF is an {\it unsupervised} outlier detection method and
tries to recognize an abnormal point in a well defined
neighborhood by evaluating the variation of the bicubic
interpolant polynomials; an outlier is then detected since
it introduces a larger variation than the neighboring
regular points do.  The method assumes some ``smoothness''
on these regular points and a certain sparsity of the
outlier distributions.  We point out that OF eliminates a
detected outlier from the set of data used in the
interpolation algorithm.  The theoretical example presented
in Section~\ref{Section_errors_in_DP} and other data
simulations not included in this paper encourage us to step
further into this direction.

Table~\ref{tab_errrrrrrrr} compares features of our methods
(ENO, OF), CRS, and Id (the extension function of Id is
identical to the raster function).  The numbers from the
accuracy column represent the maximal degree of polynomials
which can be exactly reconstructed.  The property of
recovering polynomials is closely related to the
approximation order of the method.  Data Recovery means the
capability of the method to supply the missing data from a
raster; an example of numerical errors in this matter is
presented in Table~\ref{tab_err}.  Row like grids can be
pictured as points being arranged along parallel lines; for
a mathematical definition, see \ref{ApA}.  The fidelity
feature refers to the capacity of the method to preserve the
qualitative shape of the function one has to reconstruct
(the fidelity of a method should not be confused with the
accuracy, see Figure~\ref{fig_ENO_filtering} for two
pictures of same accuracy but different shapes); the Id
extension function keeps the shape of the point
distribution.  A ranking of the methods from the speed point
of view is given in the last column, 1 for the fastest.
\begin{table}[htbp!]
  \centering
  \caption{The main features of the
    interpolation methods discussed in this paper.}
  \begin{tabular}{p{0.1\linewidth} p{0.12\linewidth}
      p{0.12\linewidth}p{0.1\linewidth} p{0.12\linewidth}
      p{0.13\linewidth} p{0.08\linewidth}}
      \hline
      Method & Accuracy & Data ~Recovery & Grid &
      Fidelity & Outlier Detection & Speed (rank)\\
      \hline
      ENO & 3 & Yes & row like & high   & No & 3\\
      OF  & 3 & Yes & row like & unknown & Yes & 3\\
      CRS & 3 & No & regular  & medium & No & 2\\
      Id  & 0 & No & any      & neutral & No & 1\\
      \hline
    \end{tabular}
    \label{tab_errrrrrrrr}
\end{table}

The water routing method presented in the
Section~\ref{Section_WaterRouting} has a physical base and
it is a simplified version of a discrete form of shallow
water equations.  One of the advantage of this approach is
the use of hexagonal raster that benefits from the higher
isotropy of hexagonal cells which allow a more suitable
modelling of the transport phenomena.  Another advantage of
this approach is that it can be incorporated in more
elaborated water flow models as a first step for a very
quick investigation of the terrain topography and the
potential water accumulation.

Further research directions include:
\begin{enumerate}
\item To develop the software in order to be compatible with
  variables having restrictions on the size and form of the
  discretization unit as a result of the observation scale
  and method.
  \begin{enumerate}
  \item to add other tessellations and irregular partitions,
    exposed as configurable user inputs;
  \item to compute the value for each polygon in the plane
    partition in two variants: at its center and by the
    average value of the intermediate function in the
    polygon, exposed as configurable user inputs;
  \item to develop an algorithm for nominal DP.
  \end{enumerate}
\item To construct a user friendly interface for the DP
  tool.
\item To develop a multiresolution analysis by using cubic
  spline wavelets in order to examine the GIS data (data
  de-noising and compression will be also considered).
\end{enumerate}

\medskip
\noindent
{\bf Remarks Concerning the Asterix Porting Data Software}\\
- The downloadable version of the Asterix Porting Data
Software available on the web is part of a larger package in
development.  The current version contains the ENO scheme,
the construction of a hexagonal raster, and the DP from a
GIS to a hexagonal raster.  In addition, a tool
for plotting the hexagonal raster is also included.\\
- This version does not contain examples of all the
theoretical and numerical results from the article.  Readers
who are interested in these examples are asked to email the
authors.\\
- The software is available under GPL license and contains
the necessary documentation for its usage.
%==========================================================

%==========================================================
\par
\medskip
\noindent
{\bf Acknowledgment} 
\medskip
\par
This work was performed within the project 50/2012 ASPABIR
\newline (www.aspabir.biogeochemistry.ro) funded by
Executive Agency for Higher Education, Research, Development
and Innovation Funding, Romania (UEFISCDI). The authors
acknowledge Florian Bodescu, who provided the digital
terrain model.  They specially thank all four anonymous
reviewers for the constructive criticism that greatly
improved the manuscript.
%==========================================================

%==========================================================
\appendix
\section{Essentially Non-Oscillating Extension Algorithm}
\label{ApA}
Here we present some details about the ENO algorithm and we
extend it to a less regular net of points.  Let
$\mathcal{D}$ be a rectangular domain
\begin{equation*}
  \mathcal{D}:=\{(x,y)\in\mathbb{R}^2\;|\; a\leq x\leq b,\;
  c\leq y\leq d\},
\end{equation*}
where $a,b,c,d$ are some arbitrarily fixed real numbers.  We
call a {\it row like grid} in $\mathcal{D}$
any set $\mathcal{N}$ of the form
\begin{equation}
  \label{annex_eno.01}
  \mathcal{N}:=\{P_{i,j}=(x^j_{i},y_j)\in\mathcal{D}\;
  |\; i=\overline{1,N_j},\; j=\overline{1,M}\},
\end{equation}
where $a<=x^j_{1}<\ldots<x^j_{N_j}<=b$ for all
$j=\overline{1,M}$ and $c<=y_1<\ldots<y_M<=d$.  Note that
the net has a variable number of knots on the lines $y=y_j$
and distance between two consecutive $y_k$ and $y_{k+1}$ is
also variable.

Let $\mathbb{L}^{\infty}(\mathcal{D})$ be the space of
bounded functions on $\mathcal{D}$ and
$\mathcal{R}:=\{g:\mathcal{N}\to \mathbb{R}\}$ the space of
reticulated functions. One defines the restriction operator
$\mathbf{R}:\mathbb{L}^{\infty}(\mathcal{D})\to \mathcal{R}$
by $\mathbf{R}(G)(P_{i,j})=G(P_{i,j})$ for all $P_{i,j}\in
\mathcal{N}$. We call $\mathbf{L}:\mathcal{R}\to
\mathbb{L}^{\infty}(\mathcal{D})$ the extension operator.

To find the extension polynomial at an interval
$I_{k}^{\xi}$, one needs to solve the minimization
problem (\ref{cubic_ext.05}) which involves the calculation
of the quantity
\begin{equation}
  \label{annex_eno.02}
  d_{p,q}^{\,k}:=
  \left\|\mathcal{P}_{(k,k+1)}^{f;(p,q)} - 
    \mathcal{Q}_k^{f}
  \right\|^2_{\mathbb{L}^{2}(I_{k}^{\xi})}.
\end{equation}
By standard calculation, we have
\begin{equation}
  \label{annex_eno.03}
  d_{p,q}^{\,k}=c\cdot\Big((\lambda
  _{p,q}^{k})^{2}+(\mu
  _{p,q}^{k})^{2}+3/2\cdot 
  \lambda _{p,q}^{k}\mu _{p,q}^{k}\Big), 
\end{equation}
where
\begin{equation}
  \label{annex_eno.04}
  \begin{array}{l}
    \delta^k_{p,q}=(\xi_{k+1}-\xi_{k})[\xi_{k};\xi_{k+1};\xi_{p};\xi_{q}]f,\\
    \lambda_{p,q}^{k}:=\displaystyle\frac{\xi_k-\xi_p}{\xi_{k+1}-\xi_k}\delta^k_{p,q}
    +[\xi_{k};\xi_{k+1};\xi_{p}]f, \\
    \mu_{p,q}^{k}:=\lambda_{p,q}^{k}+\delta^k_{p,q}.
  \end{array}
\end{equation}
and $c$ is a constant independent of the knots $p$ and $q$.
Using (\ref{annex_eno.03}) and (\ref{annex_eno.04}), the
minimization step in the algorithm (\ref{cubic_ext.05}) can
be reformulated as:
\\
\begin{minipage}{\linewidth}
  \centering \vspace*{1em} \hrule
  \begin{equation}
    \label{annex_eno.05}
    \begin{array}{l}
      \textnormal{Find } i,j \textnormal{ s.t. }
      \left\{ (\xi_{i},\xi_{j}) \right\} = 
      \operatorname*{\arg\,\min}\limits_{\xi_p,\xi_q} d_{p,q}^{\,k},\\
      \textnormal{with the pair } (p,q)
      \in\{(k-2,k-1),(k-1,k+2),(k+2,k+3)\}.
    \end{array}
  \end{equation}
  \hrule \medskip
\end{minipage}
Let $\mathbf{l}_{\xi}$ stand for the 1D continuous
extension operator from the space of reticulated functions
to the space of continuous functions.  The algorithm to
evaluate the 1D extension operator $\mathbf{l}_{\xi}$ read
as:
\\
\begin{minipage}{\linewidth}
  \centering \vspace*{1em} \hrule \medskip
  {\bf Algorithm 1.} The 1D ENO Algorithm\\
  \medskip
  % \hrule
  % \caption{The 1D ENO Algorithm}
  % \label{algorithm_1D_ENO}
  \begin{tabular}{l}
    {\bf Data Input}: $\{\xi_i\}_{i=\overline{1,N}}$,
    $f:\{\xi_i\}_{i=\overline{1,N}}\to\mathbb{R}$; $\xi$.\\
    {\bf Data Output}: $\mathbf{l}_{\xi}(f)(\xi)$.
  \end{tabular}
  \begin{enumerate}%[{\it Step} 1:]
  \item Find $I_k^{\xi}$ such that $\xi\in I_k^{\xi}$.

  \item Define $S_k^{\xi}$ of the form (\ref{Sk}).

  \item Find the knots $\xi_{i},\xi_{j}$ using
    (\ref{annex_eno.05}).

  \item Set $\mathbf{l}_{\xi}(f)(\xi) :=
    \mathcal{P}_{(k,k+1)}^{f;(i,j)}(\xi)$.
  \end{enumerate}
  \medskip \hrule \medskip
\end{minipage}

Note that the extension 1D ENO algorithm can also be applied
for the points $\xi$ outside the interval $[\xi_1,\xi_{N}]$,
using the following formula
\begin{equation}
  \label{annex_eno.06}
  \mathbf{l}_{\xi}(f)(\xi)=\left\{
    \begin{array}{lll}
      \mathcal{P}_{(1,2)}^{f;(3,4)}(\xi),& {\rm for}&
      \xi<\xi_1\\
      & &\\
      \mathcal{P}_{(N-1,N)}^{f;(N-3,N-2)}(\xi),& {\rm for}& \xi>\xi_N
    \end{array}
  \right. .
\end{equation}
Now, having the 1D ENO scheme (\ref{annex_eno.05}) and
(\ref{annex_eno.06}), one can set up the algorithm to define
the 2D extension operator $\mathbf{L}$.
\\
\begin{minipage}{\linewidth}
  \vspace*{1em} \centering \hrule \medskip {\bf Algorithm
    2.} The 2D ENO Algorithm \medskip
  % \hrule
  % \label{algorithm_2D_ENO}

  \begin{tabular}{l}
    {\bf  Data Input}:
    $\mathcal{D},\mathcal{N},g:\mathcal{N}\to\mathbb{R};(x,y)\in\mathcal{D}$\\
    {\bf  Data Output}: $\mathbf{L}(g)(x,y)$.
  \end{tabular}
  \begin{enumerate}%[{\it Step} 1:]
  \item Find $I_k^y$ such that $y\in I_k^y$.

  \item Define $S_k^y$ of the form (\ref{Sk}).

  \item For each $m$ s.t. $y_m\in S_k^y$, using Algorithm 1, calculate \\
    \hspace{6em} $f_m(x):=\mathbf{l}_x(g(\cdot,y_m))(x)$

  \item Set
    $\mathbf{L}(g)(x,y):=\mathbf{l}_y(f(x,\cdot)){\Big|_{I_k^y}}(y)$.
  \end{enumerate}
  \medskip \hrule \medskip
\end{minipage}

In this algorithm, $\mathbf{l}_x$ and $\mathbf{l}_y$ denotes
the 1D extension operators with respect to $Ox$ and $Oy$
knots, respectively.  The notation
$\mathbf{l}_x(g(\cdot,y_m))$ from Step 3 reads as follows:
the extension operator $\mathbf{l}_x$ acts on the
reticulated function $g(\cdot,y_m)$ having the values
$g(x_i^m,y_m)$ on the $Ox$ knots $\{x_i^m\}_{i=1,N_m}$.
Also, $f(x,\cdot)$ from Step 4 represents the reticulated
function having the values $f(x,y_m)=f_m(x)$ on the $Oy$
knots $y_m\in S_k^y$.

\begin{remark}
  Essentially, the 2D ENO extension operator $\mathbf{L}$ is
  given by
  \begin{equation*}
    \mathbf{L}(g) := (\mathbf{l}_y \circ \mathbf{l}_x)(g).
  \end{equation*}
  For the 2D ENO extension function the following properties
  hold:
  \begin{enumerate}
  \item $\mathbf{L}(g)(P_{ij})=g(P_{ij})$, for all
    $g\in\mathcal{R}$.
 
  \item $\mathbf{L}(G)$=G, for all $G\in\pi_{3,3}$

  \item $\mathbf{L}(g)$ is always continuous with respect to
    $y$ and continuous with respect to $x$ except for a
    finite number of points.

  \item If when on Step 4 of Algorithm 2 the problem {\rm
      (\ref{annex_eno.05})} does not have a unique solution
    for a particular $x$, then $(x, y)$ is possibly a
    discontinuity point of the extension function
    $\tilde{g}$. Otherwise, $\tilde{g}$ is locally
    continuous at $(x,y)$.
  \end{enumerate}
  \label{remark1}
\end{remark}

\begin{remark}
  Algorithm 1 and Algorithm 2 can be easily adapted for the
  OF interpolation method.
\end{remark}
%==========================================================

%==========================================================
\section{Details on test rasters from
  Section~\ref{Section_densit_behav}}
\label{ApB}
The test rasters we used in
Section~\ref{Section_densit_behav} are constructed as it
follows.

\begin{figure}[htbp!]
    \centering
    \begin{minipage}[t]{0.3\linewidth}
      \centering
      \vspace*{1.5cm}
      \begin{tabular}{ccc}
        \hline
        $\alpha$ & $m$ & $n$\\
        \hline
        $1$ & $3$ & $3$\\
        $2$ & $4$ & $3$\\
        $3$ & $5$ & $3$\\
        $4$ & $5$ & $4$\\
        $5$ & $5$ & $5$\\
        \hline
      \end{tabular}
    \end{minipage}
    \begin{minipage}[t]{0.65\linewidth}
      \centering \vspace*{0cm}
      \includegraphics[width=0.9\linewidth]{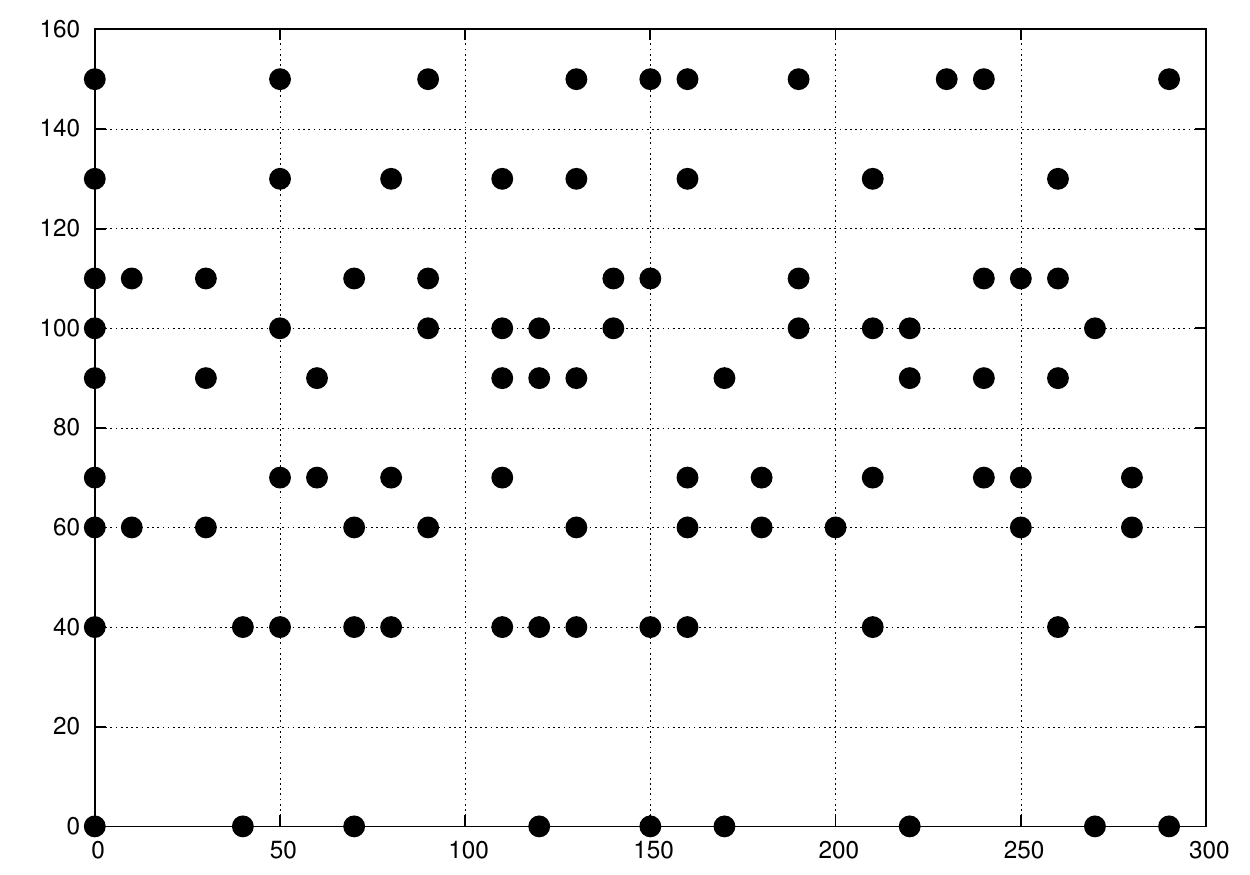}
    \end{minipage}
    %\multirow{-5}{*}{\includegraphics[width=0.6\linewidth]{random_raster_example-eps-converted-to.pdf}}
    \caption{The correspondence between the test raster
      index $\alpha$ and the parameters $m$, $n$, and a
      snapshot example (the lower left corner) of a point
      distribution in the grid for $\alpha=5$ and
      $\delta=10$ m.}
    \label{fig_test_raster_exm}
  \end{figure}

Let $\delta$ be the cell size of the square basis raster,
and $m$, $n$ some positive integers.  In order to construct
a square test raster, we randomly eliminate rows of cells
except the top and bottom rows, such that the distance
between any two consecutive remaining rows does not exceed
$m\delta$.  Then, for each of the remaining rows, we
randomly eliminate cells except the first and last ones,
such that the distance between any two consecutive remaining
cells does not exceed $n\delta$.  The Figure
\ref{fig_test_raster_exm} gives the correspondence between
the test index $\alpha$ and the parameters $m$, $n$, and an
example of a possible point distribution of the kept data in
the grid for such test.

%==========================================================

%==========================================================
%\bibliographyieeetrl2-names}
\bibliographystyle{plain}
\bibliography{porting_data}
%==================================================ieeetr\end{document}

\end{document}